\documentclass[manuscript]{acmart}
\usepackage{float}
\usepackage{subfigure}
\AtBeginDocument{%
  \providecommand\BibTeX{{%
    \normalfont B\kern-0.5em{\scshape i\kern-0.25em b}\kern-0.8em\TeX}}}
\usepackage{tabularx}
\usepackage{bbding}
\usepackage{graphicx}
\usepackage{geometry}
\usepackage{subfigure}
\usepackage{amsmath}
\usepackage{makecell}
\usepackage{float}
\usepackage{color}
\usepackage{booktabs}
\usepackage{caption}
\usepackage{tabu}
\usepackage{hyperref} % No additional options needed

\author{Yixin Chen}
\affiliation{
\institution{University of Washington}
\city{Seattle}
\country{US}}

\author{Yue Fu}
\affiliation{
\institution{University of Washington}
\city{Seattle}
\country{US}}

\author{Zeya Chen}
\affiliation{
\institution{Illinois Institute of Technology}
\city{Chicago}
\country{US}}

\author{Jenny Radesky}
\affiliation{
\institution{University of Michigan}
\city{Ann Arbor}
\country{US}}

\author{Alexis Hiniker}
\affiliation{
\institution{University of Washington}
\city{Seattle}
\country{US}}

\definecolor{DarkGreen}{HTML}{008000}

\begin{document}
%%\citestyle{authoryear}
\title[Engagement-Prolonging Design]{The Engagement-Prolonging Designs Teens Encounter on Very Large Online Platforms}

\begin{abstract}
In the attention economy, online platforms are incentivized to design products that maximize user engagement, even when such practices conflict with users' best interests. We conducted a structured content analysis of all Very Large Online Platforms (VLOPs) to identify the designs these influential apps and sites use to capture attention and extend engagement. Specifically, we conducted this analysis posing as a teenager to identify the designs that young people are exposed to. We find that VLOPs use four strategies to extend teens' use: pressuring, enticing, trapping, and lulling them into spending more time online. We report on a hierarchical taxonomy organizing the 63 designs that fall under these categories. Applying this taxonomy to all 17 VLOPs, we identify 583 instances of engagement-prolonging designs, with social media platforms using twice as many as other VLOPs. We present three vignettes illustrating how these designs reinforce one another in practice. We further contribute a graphical dataset of videos illustrating these features in the wild.
\end{abstract}

%%
%% The code below is copied from, generated by the tool at http://dl.acm.org/ccs.cfm.

%%
%% Keywords.

% \centering
% \subfigure[]{
% \includegraphics[width=0.305\textwidth, trim= 0 0 40 3]{figs/cover1.JPG}\label{fig1a}
% }\hspace{1mm}
% \subfigure[]{
% \includegraphics[width=0.305\textwidth, trim= 0 0 50 0]{figs/cover2.JPG}\label{fig1b}
% }\hspace{1mm}
% \subfigure[]{
% \includegraphics[width=0.315\textwidth, trim= 0 0 50 0]{figs/cover3.JPG}
% \label{fig:01}
% }
% \caption{Caption}
% \Description{Caption}

\maketitle

``\textit{How we spend our days is, of course, how we spend our lives.}'' ---Annie Dillard, \textit{The Writing Life}~\cite{dillard1989writing}

\section{Introduction}\label{sec:Introduction}
\begin{figure}[htbp]
    \centering
  \includegraphics[width=0.4\textwidth]{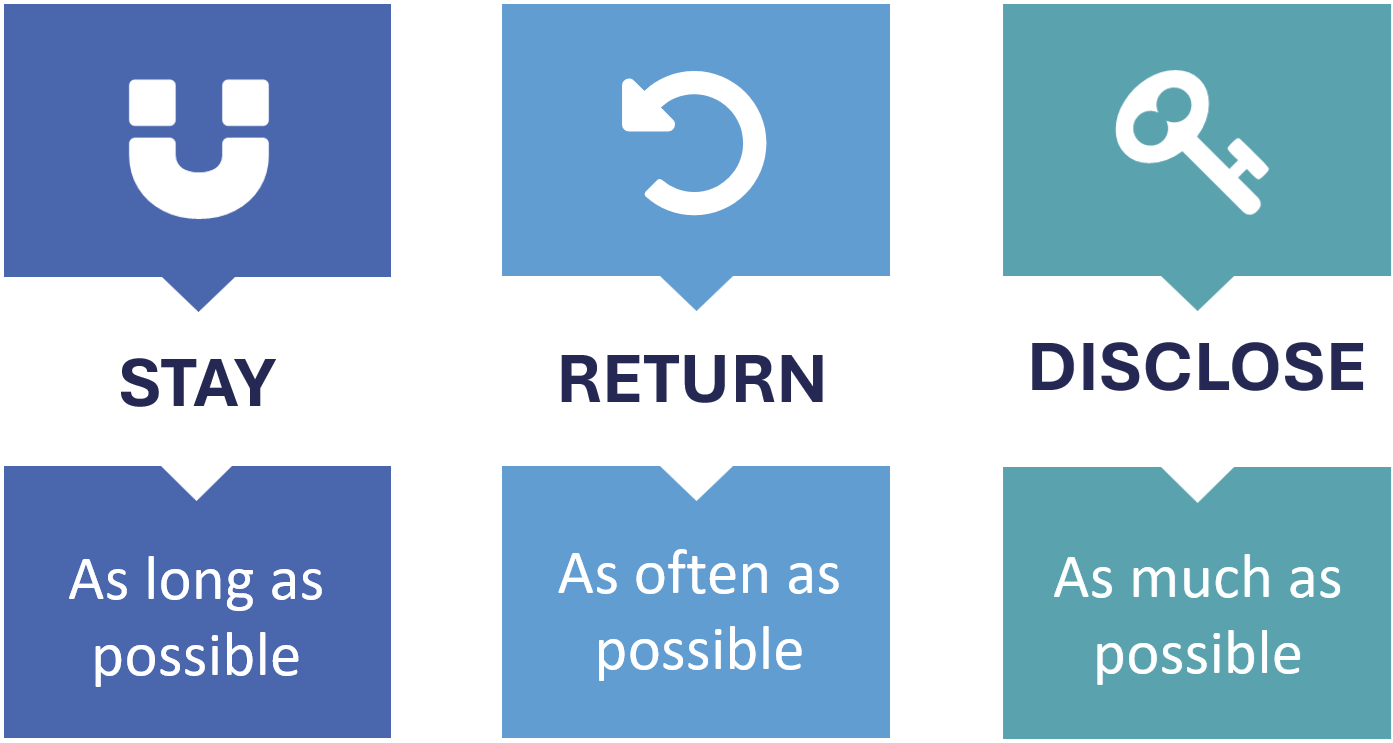}
  \caption{Attention economy incentives. Advertising-based businesses are most successful when users spend as much time as possible with the product. This increased time-on-task gives the platform the ability to serve additional advertisements and to profile the user to make these advertisements more tailored.}
  \label{fig:attentionEconomy}
\end{figure}

In the attention economy~\cite{davenport2001attention}, online platforms compete for their users' time and attention~\cite{kane2019attention}. They are incentivized to try to: 1) draw people in as often as possible, 2) keep them engaged for as long as possible, and 3) encourage them to disclose as much as possible about their habits and preferences while they are there (see Figure~\ref{fig:attentionEconomy}). This combined approach maximizes the platform's ability to serve advertisements and to make these advertisements as personalized as possible ~\cite{lambrecht2013does}. To guide a product toward this desired end-state, companies use engagement metrics like time-on-task, click-through rates, platform traffic, and session duration as primary indicators of their product’s success. They employ extensive A/B testing to engineer interfaces that lead to increases in these metrics~\cite{richards2024against}. We refer to the designs that extend the time people spend on the platform, heighten the frequency with which they interact with the platform, and deepen their disclosures to the platform and third parties, as \textit{engagement-prolonging designs} (EPDs).
%We refer to the resulting interface decisions as \textit{extended-use designs} (EUDs). 

EPDs do not always align with users’ best interests. Hartford and Stein explain that digital platforms' aggressive---and often effective---efforts to co-opt users' attention reduce people's autonomy, displace alternative activities they value, and incrementally reshape their life experiences~\cite{hartford2022attentional}. Richards and Hartzog argue that the practice of optimizing for user engagement is a form of exploitation that has long flown under the radar and deserves greater scrutiny, calling it a ``\textit{distinct and dangerous concept that should be specifically regulated}''~\cite{richards2024against}.

This claim is reflective of an increasing societal and academic interest in understanding and curtailing the practice of designing to prolong engagement~\cite{mildner2023engaging, monge2023defining, monge2022towards, docherty2022re, lukoff2021design}. And these concerns are heightened with respect to platforms' effects on youth~\cite{ hhs_youth_mental_health_social_media, woods2016sleepyteens, viner2019roles, mcculloh2023fragile}. Relative to adults, children and teens have less inhibitory control~\cite{casey2008adolescent, posner2000developing}, making them more vulnerable to manipulative designs that erode self-control. They also more likely to adopt new technologies, relative to the adults in their lives~\cite{lee2007technology}, and their social worlds are more likely to have important online dimensions~\cite{boyd2014s}, making it essential to create respectful online spaces where young people are free from manipulation by powerful companies seeking to monetize their attention.

In both North America and Europe, law suits (e.g.,~\cite{NYAG2023}), proposed legislation (e.g.,~\cite{NYGov2023}), and active legislation (e.g.,~\cite{flsenate_bill_3}) all reflect frustration with companies' attempts to prolong teenagers' engagement with their products. The European Parliament’s Internal Market and Consumer Protection Committee recently voted in favor of developing new regulations to prohibit EPDs, with particular attention to designs that affect minors. And in the United States, the U.S White House Interagency Kids Online Health and Safety Taskforce released design recommendations that call for reducing and removing features that encourage excessive or problematic use by youth~\cite{samhsa2021}.

A growing body of academic literature seeks to characterize EPDs and their consequences. For example, Monge Roffarello and colleagues report on a meta-analysis of existing literature, surfacing 11 different UI patterns that use deceptive techniques to capture users' attention~\cite{monge2023defining}. Flayelle and colleagues demonstrate that UI designs can systematically both: 1) exploit reflexive, automatic responses and 2) shape users' conscious decision-making, to encourage habitual usage patterns~\cite{flayelle2023taxonomy}. Gray and colleagues wrangle 10 taxonomies of 262 manipulative design patterns---including many EPDs---into a hierarchical ecology described by a closed grammar~\cite{gray2024ontology}. These prior efforts both reflect the HCI community's interest in characterizing this design space and demonstrate the complexity, diversity, and ubiquity of EPDs.
 
In this paper, we contribute to these ongoing efforts in two ways. First, we seek to characterize the extent to which young people in particular encounter EPDs. The European Commission~\footnote{The governmental body in the European Union that proposes and enforces legislation} has defined ``Very Large Online Platforms,'' or ``VLOPs,'' as those which have at least 45 million monthly active users in the European Union~\cite{dsa_vlops}. The designs adopted by VLOPs are poised to be the ones to which users are most exposed, given the broad reach of these platforms and the fact that many smaller companies copy patterns established by VLOPs. The platforms with the widest adoption in the EU~\cite{dsa_vlops_designation} are largely the ones with the widest adoption globally~\cite{statista_social_networks} and also the platforms which have the largest reach with teens~\cite{anderson2023teens}. Thus, we conducted a systematic evaluation of EPDs in VLOP interfaces as a way to identify the designs with the widest global impact on teen users.%measure and characterize teens' exposure to the extended-use designs . 

Second, we sought to examine these common patterns for themes in the strategies behind them. Prior work describes and clusters many discrete design patterns which have---and will likely continue to---change with time. As of yet, these patterns are not organized by the deeper strategies which motivate them and are more likely to persist even as design techniques change, providing a more stable target for regulatory action. Specifically we asked:
\begin{itemize}
\item\textbf{RQ1:} \textit{What EPDs do VLOPs present to teen users?} 
\item\textbf{RQ2:} \textit{How common are these patterns and features across VLOPs?}
\item\textbf{RQ3:} \textit{What strategies underlie these designs?} 
\end{itemize}

\noindent To investigate these questions, we conducted a two-part evaluation of all $N=16$ VLOP apps and $N=17$ VLOP websites to identify, characterize, and quantify the EPDs present on these platforms. We created accounts as teen users located in Europe to ensure the designs we encountered were ones that would be shown to teens. We conducted a thematic analysis of these interfaces to create a catalog of the EPDs they employed and then conducted a structured content analysis to more precisely understand their prevalence across this set of influential platforms. 

We find that VLOPs use four broad strategies to promote extended use. These include designs that: \textit{entice} users into engaging (i.e., using pleasure-, curiosity-, or reward-driven mechanisms to encourage a usage session), \textit{pressure} users into engaging (i.e., creating social or other obligations that encourage a usage session), \textit{lull} users into mindlessly continuing their interactions (i.e., using low-friction and autonomy-reducing designs that extend their usage session), and \textit{trap} users by erecting barriers to disengaging (i.e., using overwhelming or obfuscating designs that reduce opportunities to discontinue their usage sessions). Platforms were most likely to use a combination of strategies. 

In this paper, we showcase exemplars of each category, describe the hierarchy of patterns we encountered, report on the prevalence of each pattern, and describe three usage vignettes that illustrate how these patterns work together to prolong engagement. We further contribute a video dataset of 583 interactive instances of the 14 patterns used by VLOPs. We hope that the examples, prevalence data, and organizational scheme presented here will enable designers, researchers, and regulators to determine whether and how platforms should encourage teen users to engage with a platform.

\section{Background}\label{sec:Background}
\subsection{Manipulative Designs}
Deceptive designs that benefit a platform at the user’s expense have proliferated across digital products~\cite{mathur2021makes}. These designs have historically been referred to as ``dark patterns,'' a term that has been increasingly been replaced with ``deceptive designs''~\cite{brignull2020deceptivedesign} or ``manipulative designs''~\cite{radesky2022prevalence} to move away from racialized language that uses ``dark'' to convey a sense of danger or maliciousness.
User Experience practitioner, Harry Brignull, was the first to call widespread attention to the practice of manipulative design. Using examples from shopping and travel websites, Brignull created a taxonomy documenting patterns that deceive users into taking actions against their own intentions or best interest  (such as the ``Bait and Switch'' patterns, in which the user intends to perform one action but is tricked by the interface into performing another, or the ``Confirmshaming'' pattern, in which the platform erects a dialog box with shaming language as a barrier to proceeding). Building on Brignull’s original collection of examples, Conti and Sobiesk created a taxonomy of malicious interface design techniques, which they defined as interfaces that manipulate, exploit, or attack users \cite{conti2010malicious}. Gray et al. later presented a broader categorization of Brignull’s taxonomy and collapsed many patterns into  broader categories, such as ``Nagging'' (repeatedly making the same request to the user) and ``Obstruction''  (preventing the user from accessing functionality) \cite{gray2018dark}. 

Many studies of manipulative designs have focused on financial harms, studying eCommerce platforms that exert purchase pressure by, for example, enrolling users in hidden subscriptions or adding items to their shopping cart without consent (e.g., ~\cite{bongard2021definitely, moser2019impulse, mathur2019dark, luguri2021shining}). Other work has shown that harms are not always monetary, and Zagal et al.’s description of manipulative designs in video games categorized these patterns into: 1) temporal manipulative designs (grinding, playing by appointment), 2) monetary manipulative designs (pay to skip, pre-delivered content, monetized rivalries), and 3) social capital-based manipulative designs (social pyramid schemes, impersonation) ~\cite{zagal2013dark}. Many other studies have shown that manipulative designs frequently compromise users' privacy or attempt to harvest data deceptively~\cite{zhang2024navigating, susser2019online, bongard2021definitely}.

 %Researchers also examine manipulative patterns in specific domains. For example, taxonomies of dark patterns in e-commerce platforms have focused on purchase pressure and deceptive practices ~\cite{moser2019impulse}, and 
 
Manipulative designs broadly, including both those that lead to financial harms and those that compromise data privacy, are already the target of regulatory action. For example, the Digital Services Act Article 25 states that ``\textit{providers of online platforms shall not design, organize or operate their online interfaces in a way that deceives or manipulates the recipients of their service or in a way that otherwise materially distorts or impairs the ability of the recipients of their service to make free and informed decisions}.''

\subsection{Manipulative Designs and Engagement} 
More recently, the conversation about manipulative designs has expanded to include platforms' attempts to prolong users' engagement. Designs that prolong engagement might seem innocuous, or even potentially beneficial and reflective of a user enjoying their experience with a product \cite{wojtowicz2024social}. However, too often, EPDs are only effective because of the extreme power imbalance between companies that seek to maximize engagement and the people who use these products \cite{hartzog2021surprising}. Recent legal scholarship explains that engagement tactics are frequently unfair, deceptive, or abusive, and disproportionately benefit the party stimulating the engagement while burdening the engaged~\cite{richards2024against}. There is a pressing need to understand these manipulative EPDs, because they undermine users’ attempts to cultivate a healthy relationship with technology and set boundaries that allow them to direct their attention as needed toward sleep, work, school, time with loved ones, time outdoors, and other activities that are conducive to emotional and physical wellbeing. 

Thus, recent scholarship in human-computer interaction has begun to characterize manipulative designs that target users' attention (in contrast to earlier studies which focused on manipulative designs targeting financial deception). Monge Roffarello and colleagues conducted a systematic literature review of 43 scholarly articles \cite{monge2023defining} that identify attention-related manipulative designs and then used this meta-analysis to create a typology of 11 common attention-capture patterns (see Table~\ref{11acp}) that: 1) exploit known psychological biases and heuristics, 2) remove the need for decision-making, 3) lead the user to lose track of goals, 4) lead to a lost sense of time and control, and 5) lead to a sense of regret about the time spent on a digital service. They divided the 11 attention-capture patterns into two categories: deceptive designs that trick the user into a false belief, such as Time Fog and Fake Social Notifications, and seductive designs that tempt the user with short-term satisfaction like Guilty Pleasure Recommendations. Similarly, Flayelle and colleagues used a theoretical framework to classify the categories of designs that could contribute to addictive online behaviors such as video gaming, online gambling, cybersexual activities, online shopping, social networking, and binge-watching ~\cite{flayelle2023taxonomy}. 

There is increasing public interest in understanding and regulating manipulative EPDs. In a recent report to the European Parliament’s Committee on Internal Market and Consumer Protection, Van Sparrentak called for further legislation to address ``addictive'' EPDs in particular. This call to action lists a wide variety of manipulative practices, including quantification of social approval, overload of sensory stimuli, limited control over interfaces, notifications demanding a user return to an app on a fixed schedule, video autoplay, and pull-to-refresh content displays~\cite{monge2023defining}. 

We build on this foundational prior work in several ways. Our study expands beyond attention capture and more holistically considers the ways in which platforms both trigger and prolong usage. We further characterize the prevalence of EPDs on the most widely used technology platforms (including those most frequently used by teens), and we document the EPDs that these platforms employ specifically when these interfaces are used by young people. Regulators also need a reproducible taxonomy for identifying common EPDs, as well as the underlying psychological mechanisms being manipulated, to move forward with policy making and enforcement. To this end, we reviewed existing taxonomies of manipulative designs and theories of online behavioral addictions, and then leveraged these theories and empirical review of VLOPs to create a novel and reproducible taxonomy. 

\begin{table}[ht]
\centering
\setlength{\tabcolsep}{16pt}
\begin{tabular}{ll}
\toprule
\textbf{Pattern Name} & \textbf{Main Context of Use} \\ \midrule
P1 - Infinite Scroll & Social media \\
P2 - Casino Pull-to-refresh & Social media on smartphones \\
P3 - Neverending Autoplay & Social media and video streaming \\
P4 - Guilty Pleasure Recommendations & Social media and video streaming \\
P5 - Disguised Ads and Recommendations & Social media \\
P6 - Recapture Notifications & Social media, video streaming, messaging \\
P7 - Playing by Appointment & Video games and social media \\
P8 - Grinding & Video games and social media \\
P9 - Attentional Roach Motel & Social media \\
P10 - Time Fog & Video streaming platforms \\
P11 - Fake Social Notifications & Video games and social media \\
\bottomrule
\end{tabular}
\caption{Manipulative Designs that Target Users' Attention as Distilled by Monge Roffarello et al.}
\label{11acp}
\end{table}

%To fill this gap, we examined the designs present on all VLOPs in hopes of identifying the EPD mechanisms that have the widest global impact, and moving beyond the scope of prior examinations.

\subsection{Adolescent Development and Engagement-Prolonging Design}
A growing body of work examines the manipulative designs that child and teen users encounter. For example, Stockman and Nottingham document the use of ``cuteness'' as a UI tactic to manipulate both child and adult users~\cite{stockman2024dark}. Other work has documented the manipulative designs in games~\cite{sousa2023dark} and apps~\cite{radesky2022prevalence} used by young children, finding that these patterns are widespread in children's products. Fitton and Read~\cite{fitton2019creating} worked with early adolescent girls to create a framework for identifying manipulative designs in free-to-play apps. An exploration of teens' mental models of manipulative designs revealed that they are aware of widespread deceptive practices but often lack a precise understanding of when and how they are being manipulated in the moment~\cite{renaud2024we}.

Understanding the manipulative designs that young people are exposed to is important, because there are theoretical reasons to expect that they may be disproportionately vulnerable to deception and pressure by platforms. Teens' reduced impulse control and heightened sensitivity to rewards and social feedback make them particularly vulnerable to pressure and distraction (digital or otherwise). During adolescence, executive functions such as impulse inhibition, flexible problem-solving, and abstract reasoning develop at different rates between teens, with most of these higher-order thinking skills reaching maturity in the mid-20s. Therefore, teens are more susceptible to digital designs that exert manipulation through curiosity and novelty/risk-seeking, social pressure, and fear of missing out, with significant inter-individual variability ~\cite{nationalacademies_social_media_impact}. 

Teens’ accounts of their experiences with technology reflect frustration with attention-economy designs; more than half of teens say that they are online ``almost constantly,'' and approximately 40\% of teens say they wish they spent less time using devices and social media~\cite{pew2018teens}. As a result, many teens report attempting to cut back on these experiences. They delete apps that are too aggressive in demanding their attention~\cite{landesman2024just} and turn their account passwords over to their friends to lock them out of platforms that they find too addictive~\cite{baumer2013limiting}. They add friction by burying apps in folders ~\cite{bohmer2013study}, setting do-not-disturb times ~\cite{natarajan2024they}, and accessing platforms through browsers. But despite their makeshift strategies to resist the design tactics of the attention economy, teens still struggle to change their habits~\cite{tran2019modeling}, suffer from device-based sleep disruption~\cite{woods2016sleepyteens, viner2019roles}, and struggle with distraction from activities they value, like homework and time spent with friends and loved ones~\cite{yardi2011social}. Parents of younger children report similar experiences and say that they frequently negotiate or argue with their child about transitioning away from media, especially media that uses engagement-prolonging designs like autoplay~\cite{davis2019everything, hiniker2016not}.

In the midst of the ongoing debate regarding social media, technology, and youth mental health, there has been increasing focus on the role of engagement-prolonging design features as a mechanism linking platform use to reduced wellbeing~\cite{samhsa_online_health_safety}. Designs such as autoplay, endless scroll, and time pressure were described by the U.S. White House Interagency Kids Online Health and Safety Task Force as warranting modifications by platforms in order to reduce compulsive and excessive use of technology. Ongoing regulatory efforts have also targeted manipulative designs in products used by children and teens. For instance, the UK's Online Safety Bill aims to make social media companies legally responsible for keeping children and young people safe online \cite{Legislation2023}. And the UK’s Information Commissioner released the ``Children's Code Design Guidance'' document, directing platforms to avoid exposing children and teens to engagement nudges or other EPDs~\cite{ICO2025}. These actions suggest widespread interest in regulating platforms' ability to manipulate young people via EPDs. Our work seeks to contribute to these efforts by characterizing and quantifying the EPDs shown to teens on the world's most commonly used platforms.

\section{Method}\label{sec:Methods}
\begin{figure}[H]
    \centering
    \includegraphics[width=1.1\linewidth]{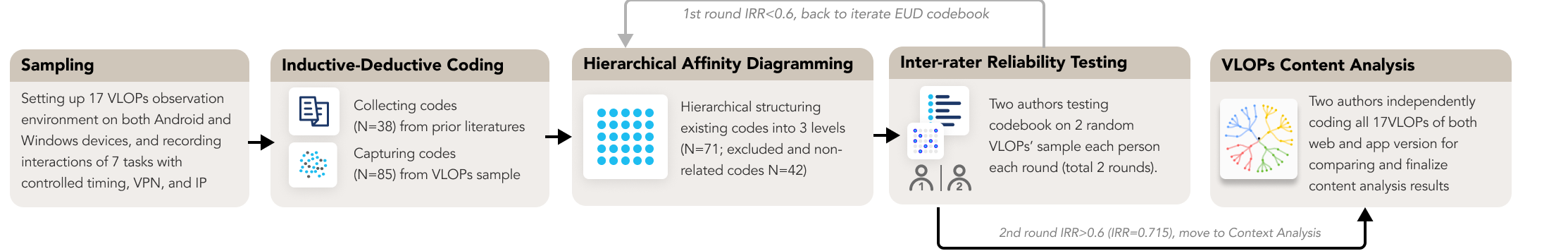}
    \caption{Overview of study procedures.}
    \label{fig:procedure}
\end{figure}

We conducted an analysis of the web and app interfaces of all 17 VLOPs identified by the Digital Services Act (DSA)\footnote{List of 17 VLOPs available on the European Commission's website: \url{https://ec.europa.eu/commission/presscorner/detail/en/IP_23_2413}} , attending specifically to the presence of EPDs. We conducted this work in two phases (see Figure~\ref{fig:procedure}); in the first phase, we performed an inductive-deductive thematic analysis to identify and categorize the EPDs present in these interfaces, examining these designs in the context of the extant literature, and organizing them by the deeper strategies that underlie them. In the second phase, we conducted a structured content analysis of all VLOP interfaces to document all instances of the EPDs we identified in phase 1.

\subsection{Sampling}
We engaged extensively with each VLOP web interface ($N=17$) and app interface ($N=16$), in each case posing as a teenage user located in the European Union to enable researchers to engage more effectively with the design space with better imagination and empathy for the user experience \cite{medler2010implications, burns1994actors, buchenau2000experience}. We adopted this method because observing these platforms requires a significant amount of time, and we believe exposing teenagers to the manipulative design patterns presented on these platforms directly for a long period is problematic.

To observe websites, we used the Chrome browser in Guest Mode on a Windows 11 laptop. To observe mobile apps, we used a factory-reset smartphone running Android 9, equipped with an active SIM card. These devices were selected because they run the most widely used operating systems in Europe~\cite{statcounter_os_share, statcounter_mobile_os_share} and therefore the operating systems that would be more commonly used by teenagers in Europe. We created two new Google accounts, which we used to access all VLOPs (we refer to these as the ``primary account,'' which was used to engage with all features, and the ``secondary account,'' which was used to enable the primary account to engage with social features that required a partner). The only exception to the above procedures was our analysis of the Apple App Store interface, which we observed using iOS devices and accounts due to the platform's restrictions.

Following the method developed by Di Geronimo and colleagues~\cite{di2020ui}, we collect screen-capture videos of each service on each modality that we could later examine in detail for evidence of EPDs. Both mobile and desktop interactions were recorded using Screen Recorder over Wi-Fi. Since platform companies may conduct A/B testing their products, the exact behaviors or interfaces we observed in a platform may not represent what all users would have encountered during the study period (the spring and summer of 2024). 
%In the studies by Moser et al., ~\cite{moser2019impulse} and Mathur et al., ~\cite{mathur2021makes}, researchers have collected screenshots of segments of pages to recognize malicious designs in e-shopping websites. In some instances, though, one can infer the presence of extended use design pattern only interacting with the artifact. That said, the UI design it self may not contain extended use design patterns, but the interaction needed to reach that interface is malicious. For this reason, 

To simulate the usage patterns of a teenage user, we created a teenager persona for our primary Google account based on prior work distilling teens' behaviors and preferences online~\cite{5rights_pathways_risk}. We used a VPN to change our IP address to a European location, and for each VLOP interface, we performed the following tasks in order, all of which were video-recorded: 
\begin{enumerate}
\item\textit{Download App and Create an Account:}
We began by downloading the app and creating a new account with our teenager persona. The primary Google account was set up with the profile of a 17-year-old girl. Additionally, we registered a secondary account with the profile of a 17-year-old boy to interact with the primary account, as detailed in step four.

\item\textit{Observe Default Settings and Interface:}
We examined the default settings, menus, and interface without making any changes to the settings or adding friends.

\item\textit{Navigate the Interface:} We next engaged with the main functions of the platform. On eCommerce platforms, we browsed the main page, systematically explored product categories and individual product pages, used search queries to find items, selected items, added them to the shopping cart, then removed them. We let the cart time out without completing a purchase. On social media platforms, we first browsed the main page, viewed posts and profiles, followed internal links to related content, and clicked on hashtags to explore similar posts. We refrained from commenting, sharing content, or directly interacting with other users, focusing solely on passive engagement. On information platforms, we browsed the main page, conducted search queries on various topics, clicked on links within articles and search results, visited related pages, and bookmarked pages for future reference.

\item\textit{Interact with Other Accounts:} If applicable, we then followed various accounts, including the secondary Google account we created, as well as brands, sponsors, celebrities, and other public accounts. The secondary account followed back and engaged with the primary account through conversation. We posted content using the primary account, with the secondary account commenting, reposting, and liking these posts.

\item\textit{Monitor Notifications and Emails:}
We then took an intentional one-week break from the platform to observe whether the platform sent notifications and emails to encourage re-engagement.

\item\textit{Re-Observe After One Week:}
After a week, we reopened the app and re-examined the platform by repeating steps three and four.

\item\textit{Disengage:} Finally, we disengaged by unfriending and blocking accounts, canceling subscriptions, and deleting both the primary and secondary accounts.
\end{enumerate}
While the screen recordings contained the visual captures of EPDs and interactions thereof, the audio served as complementary material, which allowed the coders to get a detailed impression of reviewers’ perception and judgement of a scene. We captured  both screenshot and video records of the the user interfaces during these exploration and complied annotated images in a FigJam ~\cite{Figma} collective virtual workspace as structured data samples.

\subsection{Content Analysis: Identifying the EPDs used by VLOPs}
%The development of our codebook served a crucial role not only in ensuring consistency for identifying and analyzing EUDs on VLOPs, but more importantly, in understanding and theorizing the nature and impacts of EUD itself. Our aim was to develop a comprehensive taxonomy for EUD that links: 1) psychological mechanisms through which the design would act on users, 2) patterns and strategies underlying these designs or driven by these psychological mechanisms, and 3) precise ways to describe and identify EUD through user interface features. The development process involved 6 stages, each contributing to our understanding of EUD.
We then conducted a two-part analysis to identify the EPDs used by VLOPs. We first performed an inductive-deductive thematic analysis to develop a taxonomy of the EPDs used by VLOPs. We then conducted a structured content analysis to apply it comprehensively across platforms. Here, we describe these stages in detail.
\begin{enumerate}
\item\textit{Inductive-Deductive Coding.}
We began our exploration with guidance from existing literature on attention-capture. In particular, we used the categories described by Radesky et al. \cite{radesky2022prevalence}, Monge Roferello et al. \cite{monge2023defining}, and Flayelle et al. \cite{flayelle2023taxonomy} as sensitizing concepts~\cite{tracy2013qualitative} during our review of VLOP interfaces. We also allowed for emergent themes in EPDs and noted designs that did not fit neatly into these pre-existing taxonomies. Four authors independently reviewed the VLOP interfaces using this inductive-deductive approach and documented potential EPDs in a collective spreadsheet while interacting with VLOPs as described above. We also shared and organized exemplar screenshots in a collaborative Figma~\cite{Figma} board. Consistent with prior work~\cite{di2020ui,gunawan2021comparative}, we made note of potential EPDs regardless of our perceptions of the designer's intent, allowing us to include EPDs that may not reflect malicious intentions. The team met regularly over several months, and with each iteration brought new examples to discuss. During discussions, the team compared and contrasted designs, discussed possible groupings, assessed design examples against existing taxonomies, and refined possible category definitions. This process generated 38 inductive and 85 deductive EPD codes and associated examples.

\vspace{2mm}
\item\textit{Hierarchical Affinity Diagramming.}
The research team then conducted more formal affinity diagramming~\cite{blandford2016qualitative} to restructure the codes and create an initial codebook. We structured our initial codebook into three levels: the psychological mechanisms through which the EPD design operates (high level), the recurring EPD design tactics (medium level), and the specific EPD design elements or interface components (low level). We added definitions and descriptions for each item at each level.

\vspace{2mm}
\item\textit{Inter-rater Reliability Testing.}
%Prior to initiating the formal investigation of EUD analysis on VLOPs, we conducted pilot testing with internal coders to assess the reliability of our codebook. This internal testing served two primary purposes: first, to evaluate the usability and validity of the codebook and its coding scheme, and second, to enhance the comprehensiveness and accuracy of the EUD taxonomy.
Two authors then independently reviewed two randomly selected VLOPs and applied the full set of codes across the interface. For each EPD, reviewers were required to note the code number of the mechanism and pattern, as well as specify the features exhibiting the EPDs (e.g., "1(a) gifting option"). The coding criteria focused on the distribution of EPDs across screenshots or interactions extracted from the recordings, rather than the frequency of EPDs occurrences. In those cases where an aspect of the interface was deemed a potential EPDs but was not present within the existing codebook, reviewers proposed and added a new code as a comment with a visual example. Any confusion about specific cases or the taxonomy itself was recorded as a comment. Inter-rater reliability (IRR) was established and measured following Lombard et al. ~\cite{lombard2002content}. The results of the first round yielded an IRR score below 0.6, indicating that the initial codebook (v1) fell below the minimum validity standard. Subsequently, the research team collaboratively compared the two internal coding results, discussed points of disagreement, unresolved questions, and confusion comments. This led to the development of codebook (v2), which included a refined definitions and names for each code, and the modification of codes. Codebook(v2) includes 71 EPD features in total. The same two researchers conducted a second round of internal testing with two different randomly chosen platforms, adhering to the same tasks with codebook (v2). The IRR score for this round was 0.715, exceeding the standard of 0.6 and indicating that the revised coding scheme and codebook (v2) were clear and interpretable enough to be the final version.

\vspace{2mm}
\item\textit{Assessing the Prevalence of EPDs across VLOPs.} Finally, two authors independently coded all recordings from all VLOP interfaces for each of the EPDs we documented. They then compared their coding results and came up with the final data.
\end{enumerate}

\begin{figure}[H]
    \centering
    \includegraphics[width=1\linewidth]{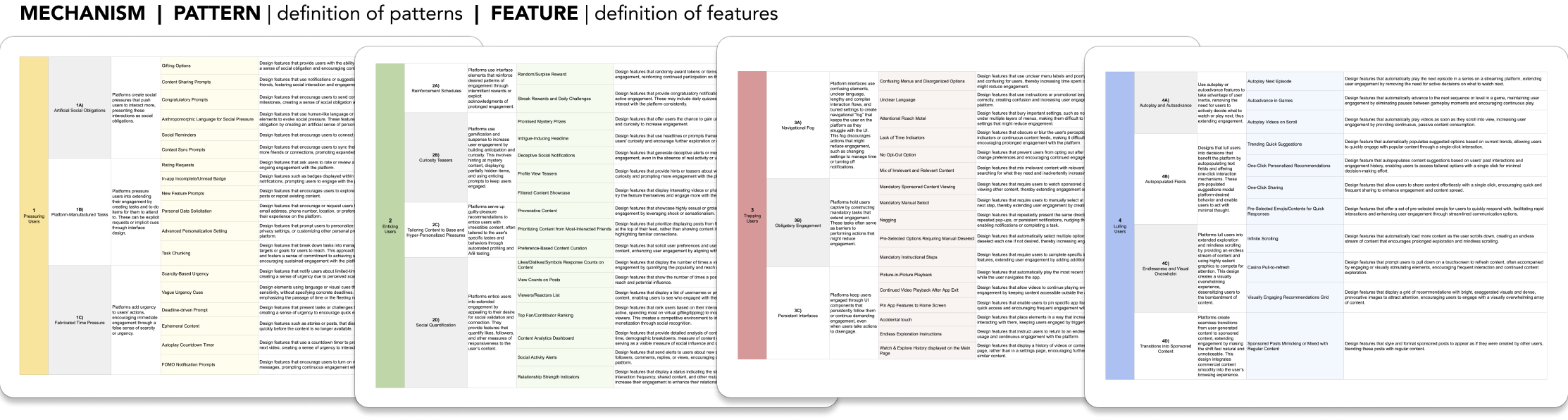}
    \Description{The image shows the final version of the EPD codebook, which includes categories and descriptions for the EPD framework.}
    \caption{Final version of the EPD codebook. More details can be found on the EPD website\protect\footnotemark.}
    \label{fig:codebook}
\end{figure}

\footnotetext{\url{https://engagementprolongingdesign.github.io/EngagementProlongingDesign/}}

\section{Results Part 1: The EPDs Used by VLOPs and the Underlying Strategies Behind Them}
We found that the EPDs that VLOPs employ fall into four broad categories, defined by the underlying manipulation strategy the design draws upon. These include: 1) artificial obligations manufactured by the platform that \textit{pressure} users into engaging, 2) reward- or pleasure-mediated mechanisms that seek to \textit{entice} users into engaging, 3) disorientating interface elements that confuse users, block navigation, or inhibit their decision-making as a way to \textit{trap} them into engaging, and 4) automated, low-friction designs that reduce user self-determination, encourage mindlessness, and \textit{lull} users into engaging. These designs can occur in combination and may overlap in the psychological mechanisms that they leverage to achieve desired effects. Many can be delivered via recapture notifications in addition to the platform’s main interface. Here, we describe each of these four approaches in more depth.

\subsection{Pressuring Users}
We found that VLOPs frequently employ designs that exert pressure on users to engage with the interface. These designs inspire a feeling that the user ought to be doing something in particular, paving the way for them to then perform the seemingly obligatory action. %Users’ time-on-task grows steadily as they work to fulfill the artificial to-do items the platform has created for them. 
We found that designs that pressure users cluster into three broad categories, including: 1) exerting peer pressure and a sense of social obligation, 2) nudging users to generate content or otherwise do work on the platform, and 3) artificial scarcity encouraging users to act quickly and engage sooner rather than later. We describe each of these categories below.

\vspace{2mm}
\noindent\textit{Artificial Social Obligations.} VLOPs translate users’ conscientiousness into extended use by presenting social obligations for the user to fulfill. This often takes the form of nudges to support or interact with other people, such as encouraging the user to congratulate a LinkedIn connection on a work anniversary or to give a virtual gift to another account on TikTok. In other instances, VLOPs encourage users to keep up with others’ content, nudge users to maintain interactions through artificial ``streaks,'' or notify users when someone has viewed their profile. They encourage users to create and post content by framing these activities as social obligations, saying things like ``\textit{Your followers want to hear from you!}'' In all of these cases, the design of the platform works together with social anxiety and reciprocity bias to pressure users into continuing to engage in interactions with other people that can only be carried out on the platform. Although notifications of these social obligations can be turned off on some VLOPs, doing so requires the user to take additional steps (e.g., LinkedIn requires four navigational steps to turn off connection-related notifications).

\begin{figure}[H]
    \centering
    \includegraphics[width=0.9\linewidth]{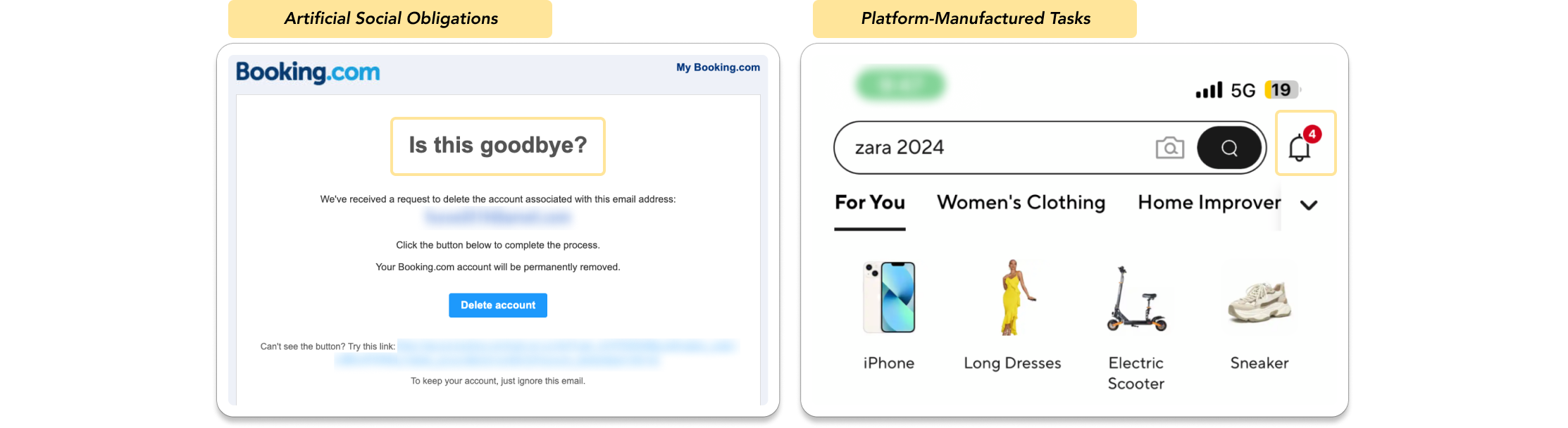}
    \caption{VLOPs pressure users by exerting peer pressure and a sense of social obligation, for example, Booking.com uses subtle anthropomorphization to emotionally manipulate a user who attempts to delete their account (left); and by nudging users to generate content or otherwise do work on the platform, for example, immediately upon sign up, an AliExpress user has multiple notifications to tend to, as indicated by the red bell in the upper right corner of the screenshot above. The salience of this design creates a sense that the user should be checking on these notifications and dealing with them, but they are often irrelevant or promoted content (right).}
    \label{fig:artificial}
\end{figure}

VLOPs also anthropomorphize their interfaces in ways that create social pressure to engage according to the platform’s interests. These anthropomorphic elements encourage the user to respond to the platform as if it were a person, creating pressure to treat the platform ``well'' (which the platform defines as extending engagement). For example, if a user attempts to delete their account on the Booking.com platform, a confirmshaming dialog asks, ``\textit{Is this goodbye?}'' (see Figure ~\ref{fig:artificial}), manufacturing a subtle guilt trip and conflating the act of preserving an account on the platform with the act of tending an interpersonal relationship.

\vspace{2mm}
\noindent\textit{Platform-Manufactured Tasks.} VLOPs also pressure users into extending their engagement by constructing tasks and to-do items for them to attend to. In some cases, these are explicit requests, for example, asking users to provide ratings, check in-app notifications, or post content. In other instances, the interface design implicitly constructs unstated tasks for the user to perform, such as clearing notification badges or completing collections.

\vspace{2mm}
\noindent\textit{Fabricated Time Pressure.}
VLOPs add urgency to users’ actions on the app or site, encouraging them to hurry back to engage. In some instances, this takes the form of artificial countdown timers or a false sense of scarcity. If a user has to ``act now,'' they must immediately engage with the interface. In other instances, VLOPs create time pressure by making content ephemeral, manufacturing fear of missing out if the user does not return to the platform in time to view it. This helps build the compulsion to return to the app just in case new content is available and helps to cement daily checking habits. VLOPs use push notifications to build both one-time urgency (e.g., time-limited offers) and ongoing urgency (e.g., vanishing stories).

\begin{figure}[H]
    \centering
  \includegraphics[width=0.9\linewidth]{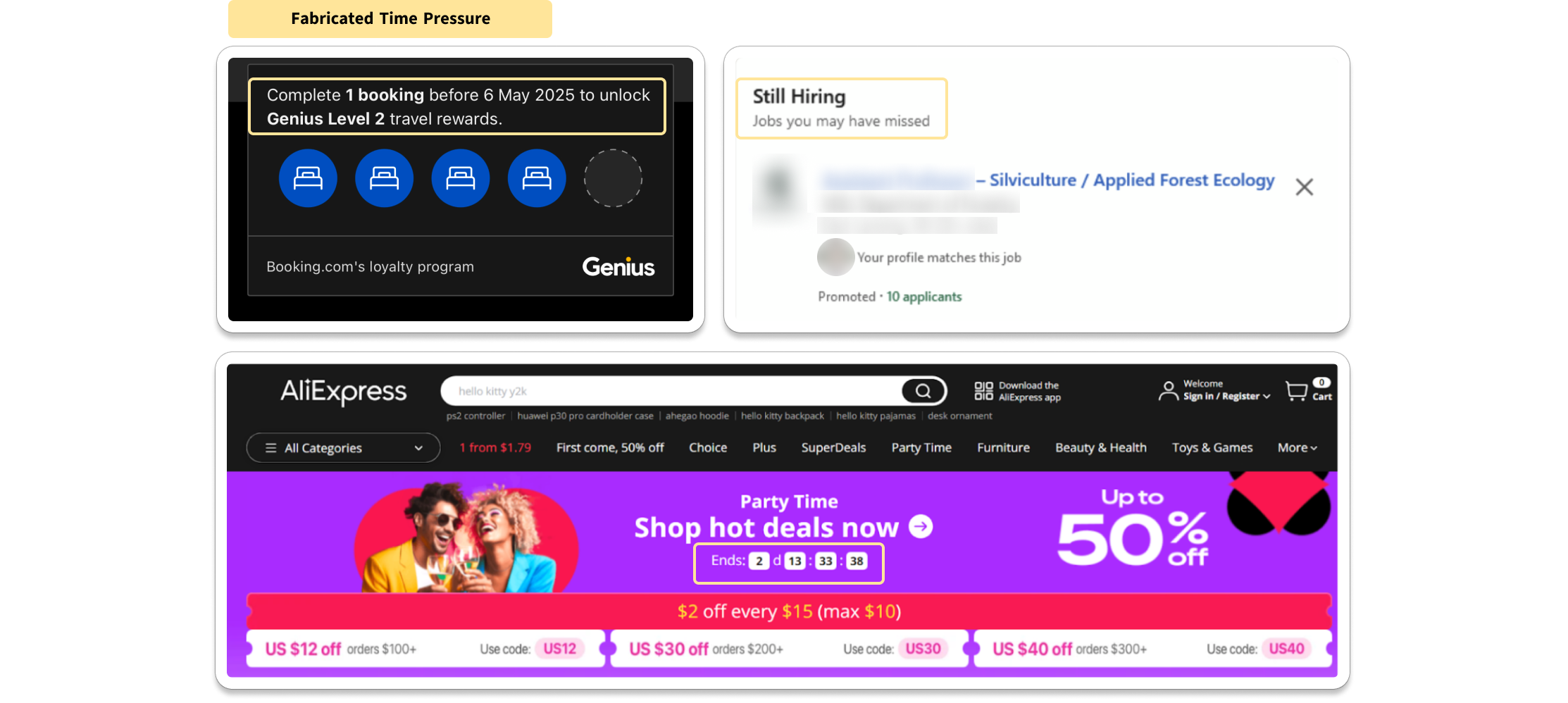}
  \caption{VLOPs cite urgent circumstances and fleeting rewards to pressure users into staying on the platform and deepening their engagement with it. For example,  if the user continues to use booking.com regularly, they will receive travel rewards (top left), LinkedIn warns the user that they may have only a brief window of time to consider a job opportunity they have missed (top right), and AliExpress offers the user a maze of endless content surrounded by timers and coupons suggesting they  must browse now or forego these savings.}
  \label{fig:interface}
\end{figure}

\subsection{Enticing Users}
Separately, we found that VLOPs encourage reward- and pleasure-seeking via algorithms that prioritize engaging content. This incentivizes content creators to cater to impulsive, subconscious preferences that generate fast-brain clicks governed by "System 1" thinking \cite{kahneman2011thinking}. This content is then organized within the interface to be maximally enticing, and when users step away from the platform, suspense-building notifications attempt to draw them back. Presenting potentially enticing content has the secondary benefit of allowing the platform to profile the user, learning their particular vulnerabilities by observing the types of clickbait---including those created by generative artificial intelligence---that are most effective in extending their engagement.

\begin{figure}[H]
    \centering
  \includegraphics[width=0.9\linewidth]{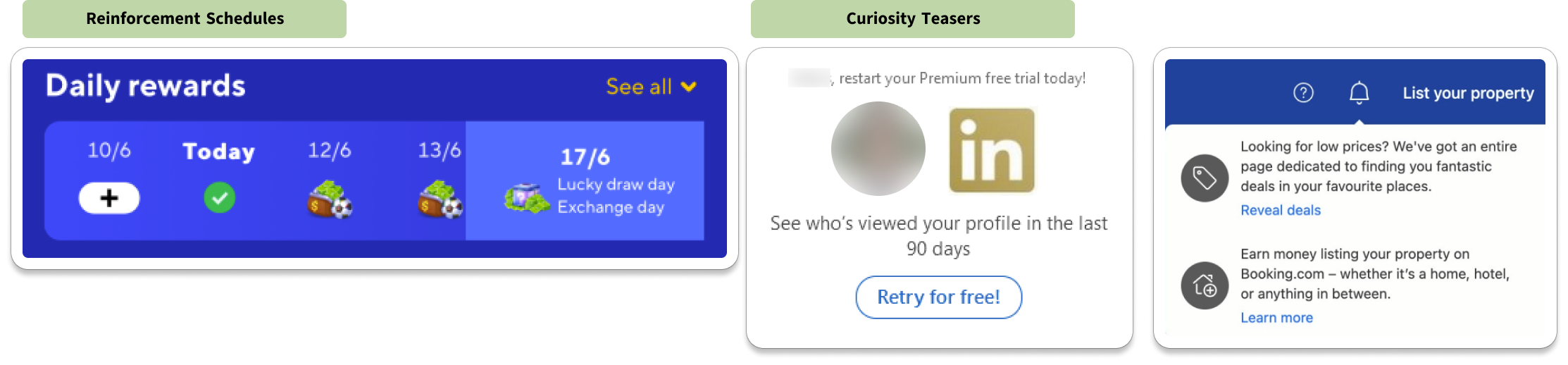}
  \caption{As users navigate through VLOPs, they are baited with promised mystery prizes as rewards for daily engagement (AliExpress, left), teasers about who is looking at their profile (LinkedIn, middle), encouraged to stay on the site to keep digging for the best deal (booking.com, right).}
  \label{fig:curiosity}
\end{figure}

\vspace{2mm}
\noindent\textit{Reinforcement Schedules.}
VLOPs frequently employ interface elements that reward and affirm users for engaging with the platform or extending their time-on-task.  In some instances, this involves offering rewards (such as free items, tokens, social rewards like ``likes,'' or pleasurable content) on intermittent or obscured reinforcement schedules, triggering dopaminergic pathways and increasing cravings to return to the app in hopes of better luck next time. In other instances, the platform reinforces prolonged engagement by rewarding it explicitly, for example, by rewarding daily logins, congratulating users on their activities within the app, or conferring badges or status as a function of engagement (Figure~\ref{fig:curiosity}, left).

\vspace{2mm}
\noindent\textit{Curiosity Teasers.}
VLOPs also leverage gamification and suspense to increase usage by building users’ anticipation. For example, VLOPs hint at mystery content by showing just a glimpse of an item, displaying it partially off-screen and encouraging users to continue to scroll to satisfy their curiosity. In other cases, VLOPs promise information rewards via irresistible prompts like ``See who’s viewed your profile'' (Figure~\ref{fig:curiosity}, middle) or the promise of better deals if they continue exploring (Figure~\ref{fig:curiosity}, right). These teasers encourage the user to continue engaging to resolve a mystery that has been artificially constructed by the platform.

In many instances, VLOPs embed these teasers into recapture notifications, extending their reach into moments when the user is engaged elsewhere. These notifications encourage new usage sessions by building suspense, promising rewards, and touting the validation of others, with messages like ``\textit{Save on something you’ll love}'' (Amazon), ``\textit{Check out some of today’s most watched reels}'' (Instagram), and ``\textit{People are reacting to [user name’s] post}'' (Facebook).

\begin{figure}[H]
    \centering
  \includegraphics[width=0.9\linewidth]{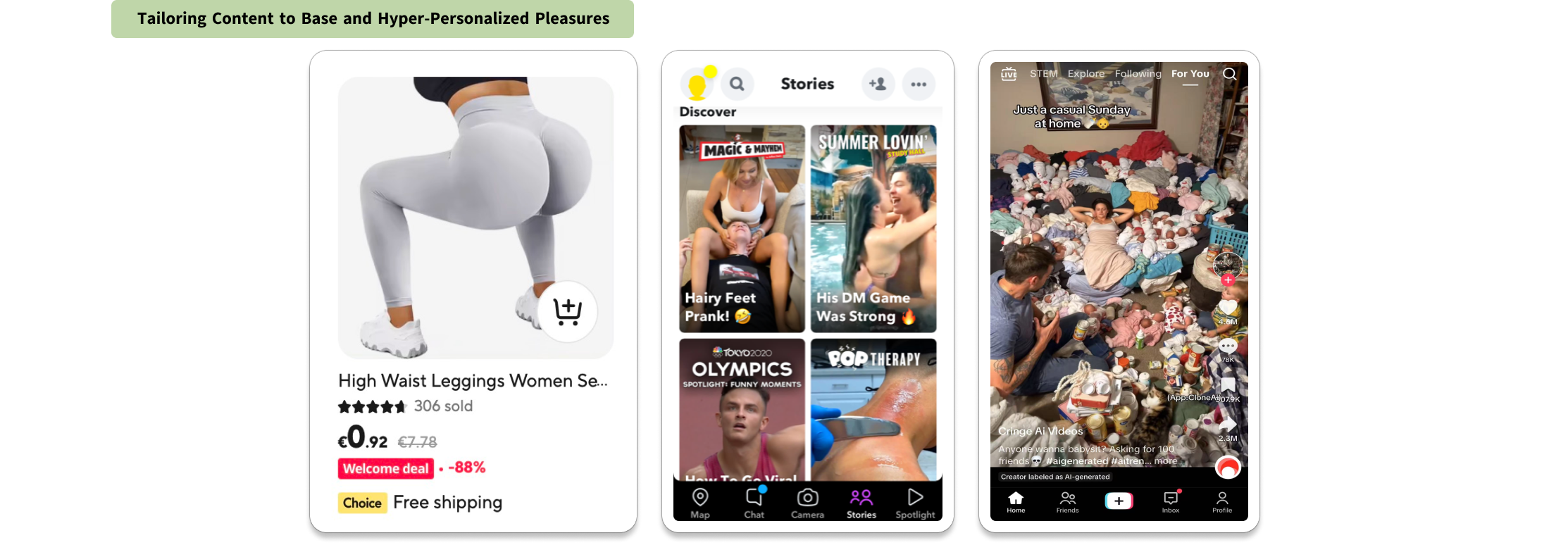}
  \caption{VLOPs attempt to extend engagement by appealing to base pleasures, enticing users with sexual, violent, or grotesque content. Left: AliExpress uses sexualized images to attract clicks and direct users to product links; Middle: Snapchat promotes violent user-challenge videos to capture attention; Right: TikTok leverages employs grotesque content to provoke reactions and encourage user interaction.}
  \label{fig:pleasure}
\end{figure}

\vspace{2mm}
\noindent\textit{Tailoring Content to Base and Hyper-Personalized Pleasures.}
Additionally, VLOPs serve up guilty-pleasure recommendations to entice users with irresistible content. In these instances, the user does not search for or seek out this content; it is proactively delivered by the platform. Hypersexual and grotesque images extend users’ time on the platform by producing a feeling of being unable to look away (see Figure~\ref{fig:pleasure}). VLOPs bombard users with guilty-pleasure content in feeds, advertisements, popups, mainline content, and sidebar suggestions.

With repeated use, we found that as a platform profiles a user and comes to know their tastes with more precision, these guilty-pleasure recommendations become more fine-tuned to the specific triggers that are likely to elicit clicks from the particular user. These are potentially less gratuitous and instead are tailored to the interests of the user (which prior has shown can increase the likelihood of the user feeling ``understood'' by the platform~\cite{gilmore2020affinity}). In practice, these curated feeds are shaped by automated A/B testing that examines which recommendations lead to the longest engagement; this testing matches content heuristics (like humor, sexiness, luxury, drama, food) to the user’s engagement behavior ~\cite{gilmore2020affinity}.

\begin{figure}[H]
    \centering
  \includegraphics[width=0.8\textwidth]{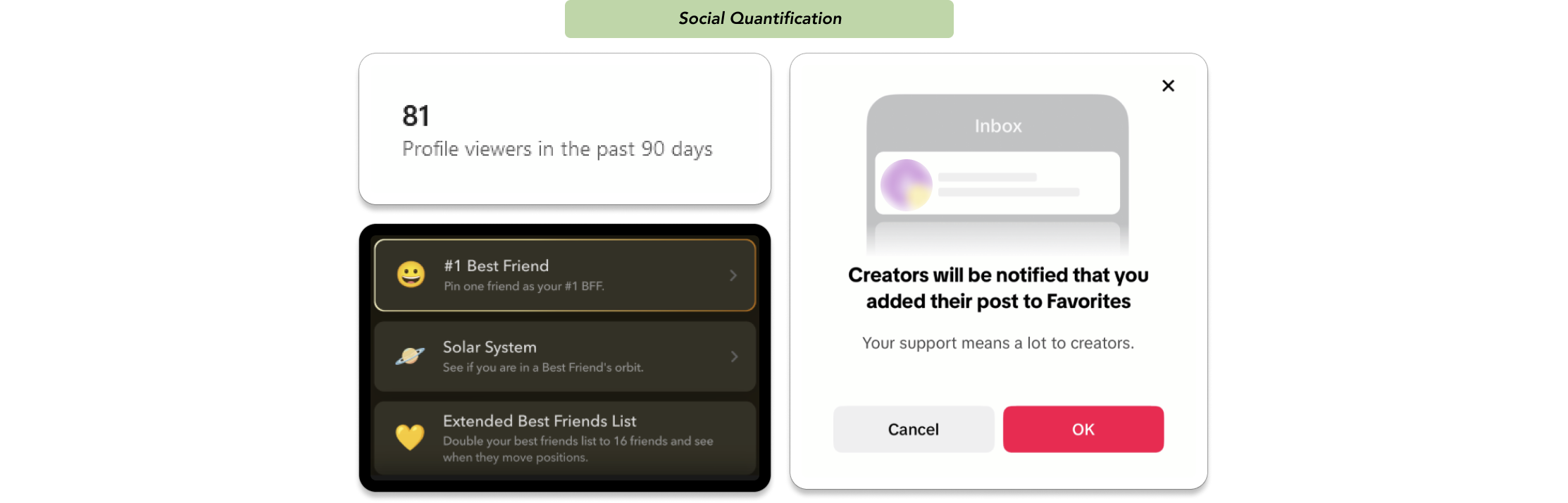}
  \caption{VLOPs create features that quantify social approval, motivating users to engage more extensively with the platform to improve these metrics. Top Left: LinkedIn quantifies profile views as a measure of the interest of others; button left: Snapchat offers features to quantify the strength of friendships; right: TikTok quantifies indicators of approval and surfaces these to the creator of the content.}
  \label{fig:social}

\end{figure}

\vspace{2mm}
\noindent\textit{Social Quantification.}
We found that, in other instances, VLOPs entice users into extended engagement by appealing to their desire for social validation and connection. They provide features that enable affirmation from others, quantifying likes, followers, and other measures of responsiveness to the user’s content (see Figure~\ref{fig:social}). This encourages the user to engage with the platform, first, by sharing content that others are likely to respond to, and second, by returning again and again to track the validation they may or may not be receiving from others. Like curiosity teasers, these social rewards are often embedded into push notifications (e.g., ``\textit{You have two new followers}''), enabling the platform to encourage engagement and redirect the user’s attention back to the platform, even when they are otherwise occupied.

\subsection{Trapping Users}
We further observed that VLOPs frequently use designs that attempt to hold the user captive. They achieve this effect via: 1)  confusing interfaces that make it difficult for the user to pursue disengagement goals, 2) obligatory activities that artificially stand between the user and their objectives on the platform, and 3) persistent content that continues to inject itself into the user’s frame of attention even after they disengage. Here, we describe each of these three types of designs. In the short term, when users encounter UI traps, they may fail to engage with the platform as they intend to. With repeated exposure to these designs, users may develop resignation about their ability to control what happens on the platform or a defensive stance towards interactions with the platform.

\vspace{2mm}
\noindent\textit{Navigational Fog.}
Platform interfaces use confusing elements, unclear language, lengthy and complex interaction flows, and buried settings to create navigational ``fog'' that keeps the user on the platform as they wrestle with the UI. Fog is selectively designed to discourage users from taking actions that might curtail their current or future engagement, like changing settings to manage time on the platform or turning off notifications. For example, Figure 9 shows the five steps needed to log out of TikTok, each an opportunity for the user to lose their way and remain on the platform. In contrast, logging into the platform is quite simple, requiring only a single step with easy-to-navigate UI (Figure 10, far right). In another instance, the user is shown a notification badge, but has to navigate through dozens of menus and submenus to discover the item that triggered the notification.

\begin{figure}[H]
    \centering
  \includegraphics [width=1\linewidth]{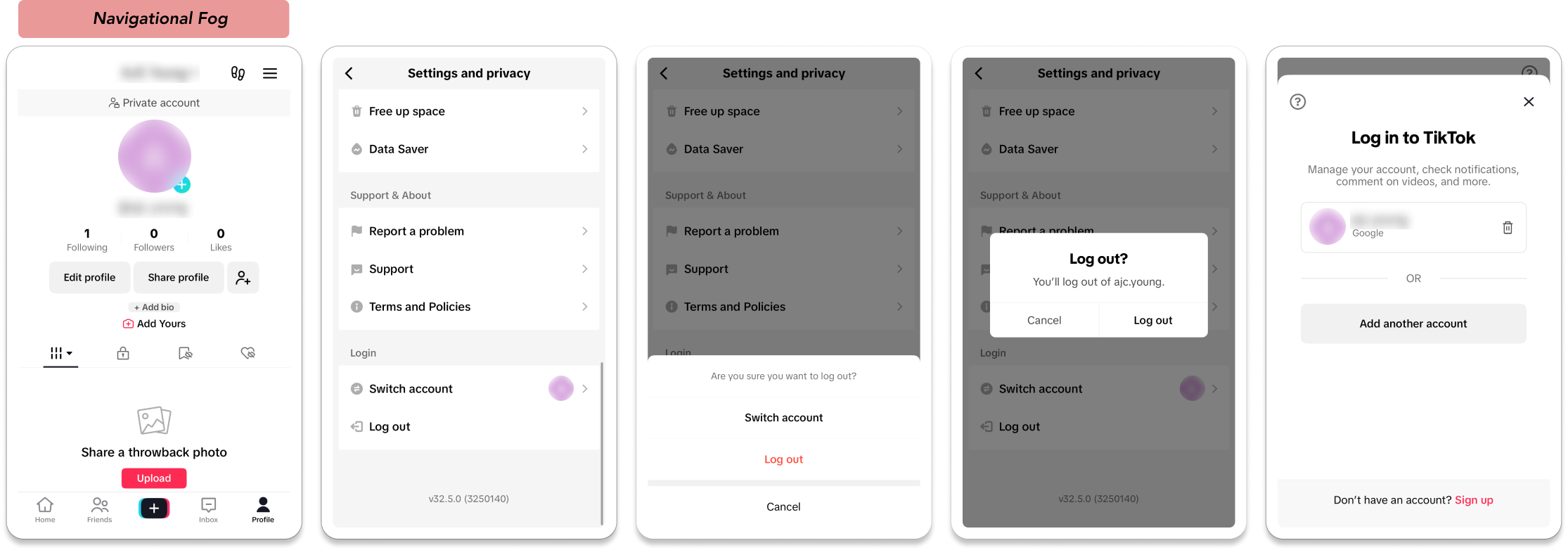}
  \caption{Navigational Fog. Logging out of TikTok requires five different steps, several of which are difficult to find. To discover how to log out, our team had to use a separate search engine to learn where to find this option.}
  \label{fig:interface}
\end{figure}

VLOPs also use confusing language, making it difficult for the user to discern what they are looking at and whether it is of interest. This disorientation has the effect of keeping users on a platform longer as they search to find content of interest or confirm that they have found the best option. Prior work documents that teens extend their engagement on social media VLOPs by wading through a ``soup'' of irrelevant content, reporting that with enough scrolling and searching they eventually stumble upon nuggets they find interesting~\cite{landesman2024just}. This wading is by design, and other work reports that VLOPs intentionally intersperse enticing content with tedious content to keep users on the hunt~\cite{meta_lawsuit2023}. Similarly, we found that VLOPs bombarded users with text that was difficult to interpret (like, ``\textit{First come, 50\% off}'' (AliExpress)), sponsored posts masquerading as user-generated content, and layers of menus and settings to dig through.

% \begin{figure}[htbp]
%     \centering
%   \includegraphics[width=0.8\textwidth]{figs/11.jpg}
%   \caption{VLOPs attempt to hold the user captive in confusing, hard-to-navigate interfaces. Left: Zalando signals to the user that there is something they need to check (via the orange notification in bottom right corner) but does not specify where this comes from, leaving the user to dig through dozens of menus and submenus. Right: the interface for AliExpress is blanketed with promotions that are difficult to decipher.}
%   \label{fig:interface}
% \end{figure}

\begin{figure}[H]
    \centering
  \includegraphics [width=0.9\linewidth]{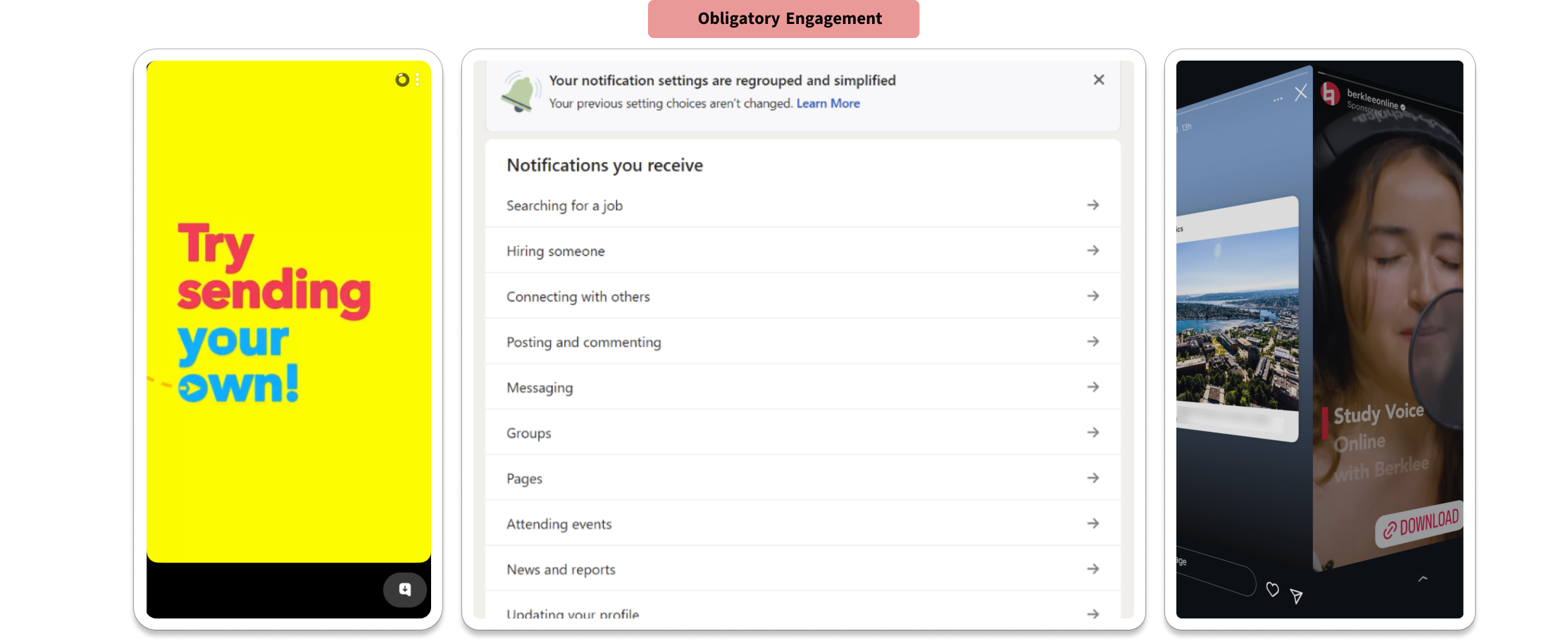}
  \caption{VLOPs construct obligatory tasks for users. Left: A video advertisement blocks the screen on Snapchat; the user cannot continue until they watch the video to completion. Middle: to turn off LinkedIn notifications, the user must individually select each subcategory of notifications to turn off. Right: Instagram inserts ads within Stories that users must watch before they can continue viewing the Stories of the accounts they follow.}
  \label{fig:obligatory}
\end{figure}

\vspace{2mm}
\noindent\textit{Obligatory Engagement.}
In other instances, VLOPs attempt to hold their users captive by constructing mandatory tasks that extend their engagement. For example, on Snapchat, users must watch video advertisements to completion before performing desired activities, and on Instagram, users must watch sponsored Stories if they want to continue viewing the Stories of the people they follow (see Figure~\ref{fig:obligatory}). At times, these obligatory tasks are erected as barriers to performing tasks that might end engagement; for example, LinkedIn does not allow users to turn off notifications holistically, and instead requires the user to manually opt out of every possible notification type.

\begin{figure}[H]
    \centering
  \includegraphics [width=0.9\linewidth]{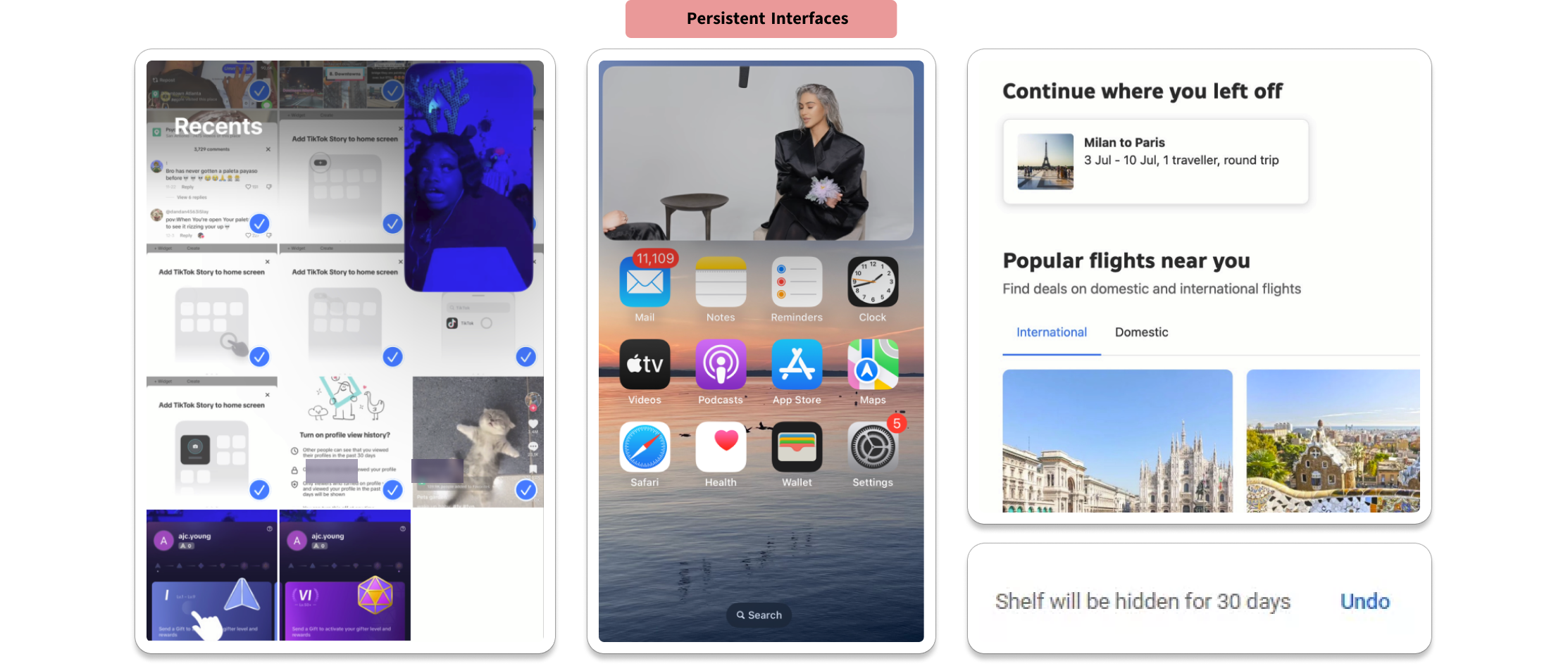}
  \caption{VLOPs badger users into continuing an engagement behavior even after they have taken steps to disengage. Left: TikTok continues playing a video even after the user navigates away from it. Middle: YouTube continues playing a video after the user leaves the app entirely. Top right: Booking.com brings a user back to past searches and nudges them to continue an engagement stream they had previously abandoned. Bottom right: YouTube allows the user to temporarily hide their attention-grabbing ``Shorts'' section but after 30 days, brings this feature back without the user’s consent.}
  \label{fig:persistent}
\end{figure}

\vspace{2mm}
\noindent\textit{Persistent Interfaces.}
Finally, VLOPs attempt to keep users engaged via UI components that aggressively follow them or continue to demand engagement, even if the user takes action to disengage. For example, even as a user leaves the main feed and navigates to other parts of the app, TikTok continues to play the most recent video in a picture-in-picture frame with no options to stop the video or mute it (Figure~\ref{fig:persistent}, left). Booking.com resurfaces searches to the user that they had previously chosen to walk away from and did not proactively save, encouraging them to ``continue where you left off'' (Figure~\ref{fig:persistent}, top right). In these and other examples, the interface clings to the user’s past engagement behavior and pushes the user to resume it, even though they have actively disengaged.

\subsection{Lulling Users}
User interfaces are often designed to make it as easy as possible to continue engaging or to explore new areas of the platform. This induces artificial mindlessness, leading the user to select easy or default options and to click and navigate thoughtlessly. Although creating ``frictionless'' experiences is an objective of UX design that reduces cognitive burden and unnecessary decisions, it can also serve the purposes of the platform at the expense of the user by taking over their decision-making and lulling them into a more passive, automatic cognitive state. We found that low-friction designs appear across VLOPs with varying degrees of transparency and user control.

\vspace{2mm}
\noindent\textit{Autoplay and Autoadvance.}
We found that VLOPs that stream videos frequently implement autoplay or autoadvance features which take advantage of the user’s inertia, remove the need to search for what to watch or play next, and attempt to take over the user’s decision-making. The end of a video or gaming sequence offers a natural stopping point for the user and invites a decision about whether to continue or end engagement; autoplay attempts to downplay this decision point and lull the user into continuing to engage. 

Similarly, we found that on most VLOPs with video content, videos autoplay as soon as they scroll into view, maximizing the chances of attracting the user’s attention and attempting to deepen the user’s engagement in moments when they might not have opted to continue engaging. These features reflect a broader pattern of platforms presuming engagement intentions on the part of the user, that is, responding as if the user has signaled an interest in engagement (even when they haven’t) and ignoring the user’s signals that they want to disengage.

\begin{figure}[H]
    \centering
  \includegraphics [width=1\linewidth]{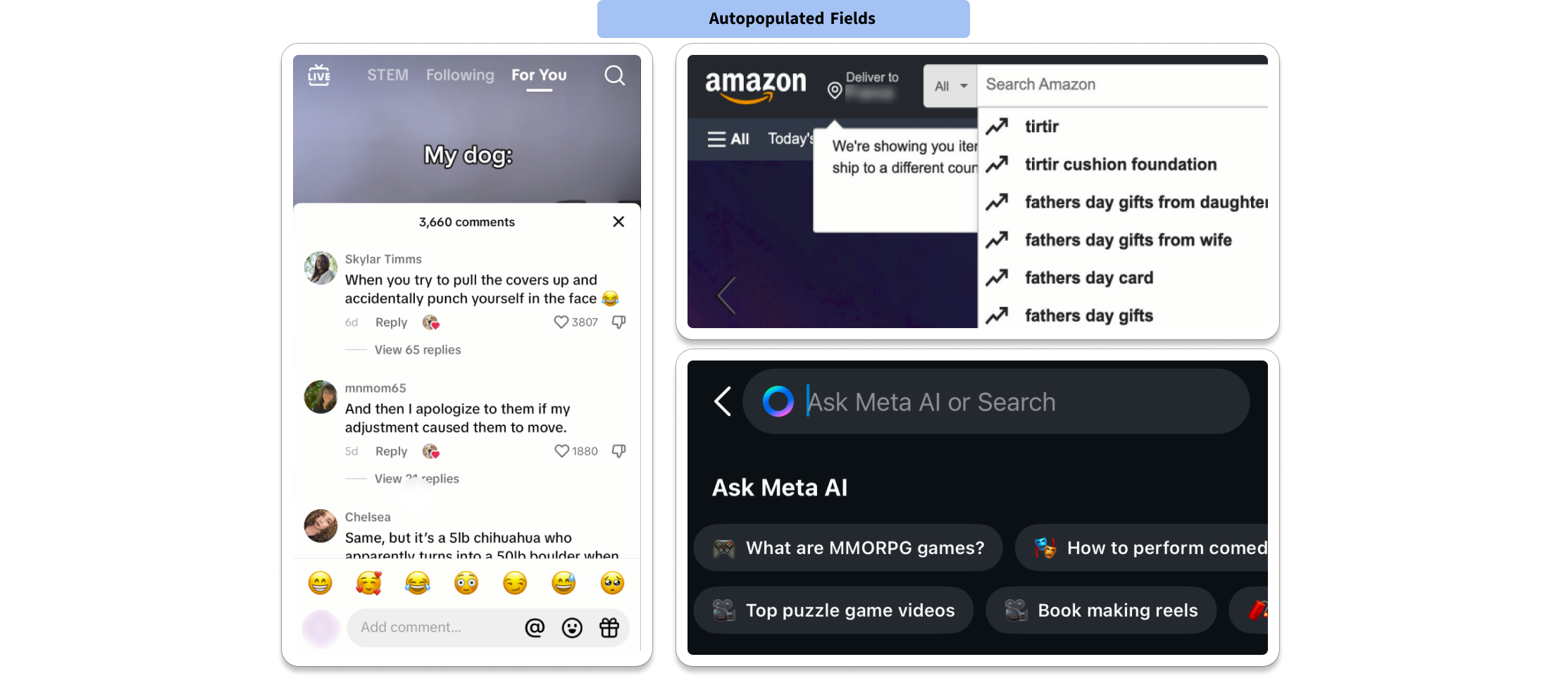}
  \caption{Autopopulated responses and searches lull the user into taking action with almost no thought required. Left: TikTok serves the user possible responses to another person. Right: Amazon (top) and Instagram (bottom) fill out search suggestions.}
  \label{fig:autofill}
\end{figure}

\vspace{2mm}
\noindent\textit{Autopopulated Fields.}
Interfaces also lull users into continued engagement by autopopulating text fields and offering one-click interaction mechanisms. Feeding the user pre-populated text both models platform-desired behavior for the user and makes it possible for the user to enact this behavior with a single tap or keystroke. For example, Instagram proactively suggests possible entertainment searches based on the user’s profile, clicking Amazon’s search bar brings up a list of suggested shopping searches, and TikTok suggests the user respond to a post by offering several pre-selected emojis to choose from (Figure~\ref{fig:autofill}).

VLOPs often implement this approach together with enticement designs, lulling users into mindlessly following curiosity teasers and engaging in other novelty-seeking behaviors on autopilot. For example, platforms autopopulate search bars with guilty-pleasure content and content labeled ``trending'' to pique users’ curiosity.

\vspace{2mm}
\noindent\textit{Endlessness and Visual Overwhelm.}
VLOPs also lull users into extended exploration and mindless scrolling by providing an endless stream of content. Feed-based platforms frequently employ an infinite scroll design, serving content indefinitely and seamlessly without ever injecting natural stopping points or giving space for the user to question whether they have seen enough. Similarly, heuristics incentivize content creators to use highly salient graphics to compete for users’ attention. This leads platforms to saturate all content area, creating a visually overwhelming tableau and desensitizing the user the experience of being bombarded with content.

Endless, visually overwhelming content is nearly ubiquitous, and for example, YouTube’s recommendations grid displays high levels of color saturation and high density of objects (coupled with content designed to leverage enticement designs, including manufactured drama, peril, and frightening images). The Google Play app store contains several screens’ worth of app thumbnails with similar visual characteristics. AliExpress contains thumbnails featuring women’s or men’s bodies, distorted pictures, or images that tap into violent cognitive biases.

\begin{figure}[H]
    \centering
  \includegraphics [width=1\linewidth]{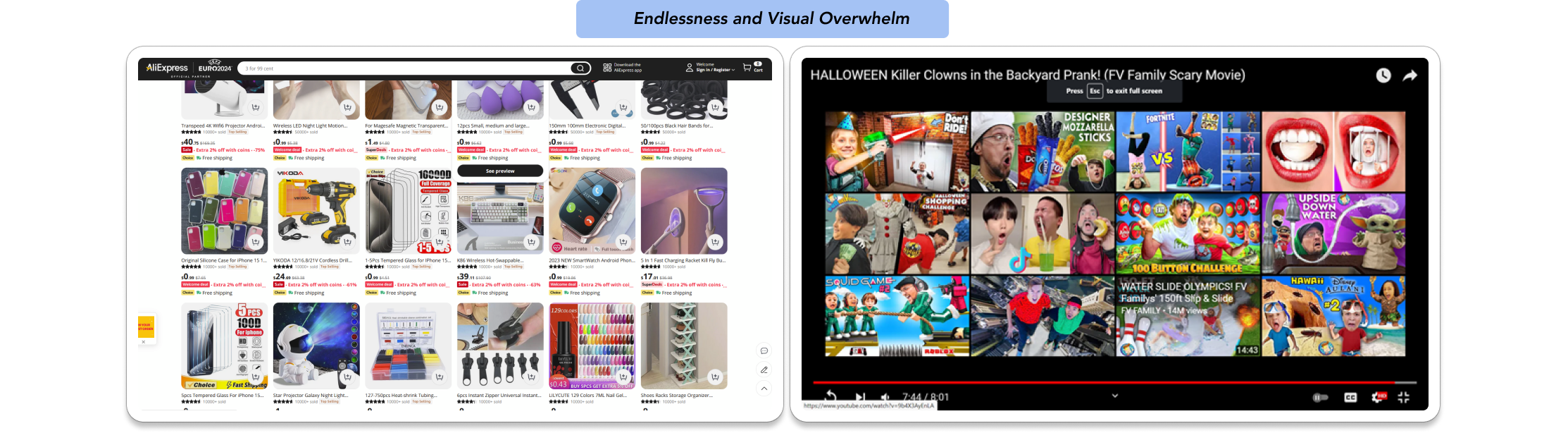}
  \caption{An infinite number of visually salient options on AliExpress (left), YouTube (right).}
  \label{fig:overwhelm}
\end{figure}

\begin{figure}[H]
    \centering
  \includegraphics [width=0.7\linewidth]{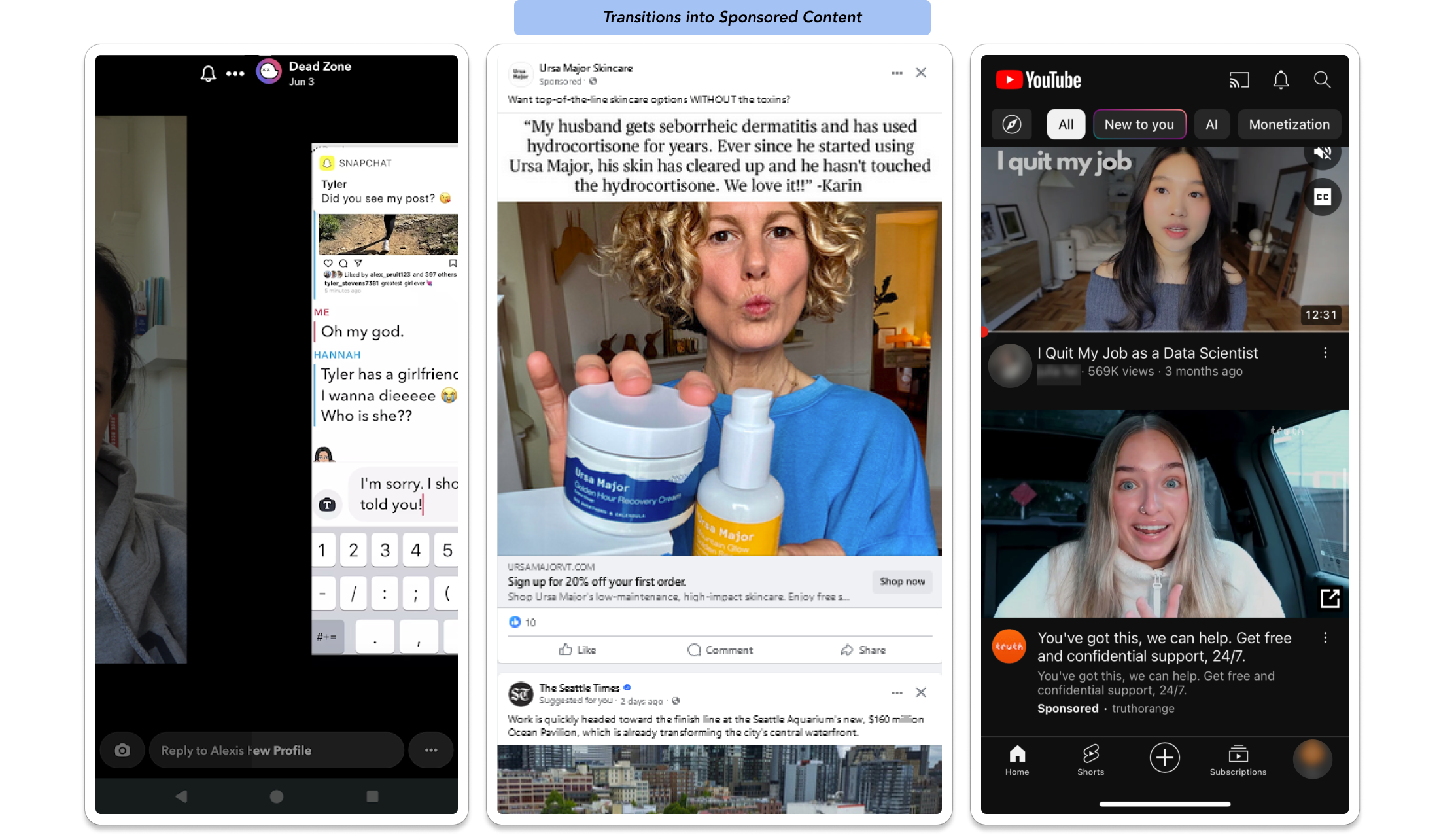}
  \caption{Platforms transition users seamlessly from interactions with friends into marketing content. Left: After browsing friends’ stories in SnapChat (a friend story shown partially offscreen to the left), swiping again transitions the user into an influencer’s story. Middle: Facebook encourages users to extend use by hunting for content in a feed of sponsored (top) and suggested (bottom) posts. Right: YouTube plays ads in the same format and style as platform content.}
  \label{fig:transitions}
\end{figure}

\vspace{2mm}
\noindent\textit{Transitions into Sponsored Content.}
Finally, platforms often pave a smooth path from user-generated content to sponsored and recommended content that the user has not sought out. This extends engagement by expanding the set of available stimuli, and platforms rely on lulling designs to make this transition feel seamless. For example, a Snapchat user can swipe left to browse through their friends’ stories; when they reach the last story, swiping again will bring up a public story from an influencer without alerting the user to the fact that they have exhausted all of the content from friends (see Figure~\ref{fig:transitions}, left). Similarly, YouTube and Facebook lull users into consuming advertising content, extending their usage time by encouraging them to hunt for interesting content and making the shift from personal to commercial content seamless and normalized.

\section{Results Part 2: Prevalence and Distribution of EPDs on VLOPs}
After distilling and organizing the patterns described above, we systematically coded all 17 VLOP websites and their 16 corresponding apps for the presence of these designs. In doing so, we both documented the specific features they used to invoke these EPD patterns and the frequency of each type of pattern across the ecosystem of VLOPs. Further supplementary material, including interactive visualizations of this data, a database of feature instances, and our codebook is available on the Engagement-Prolonging Design website\footnote{\url{https://engagementprolongingdesign.github.io/EngagementProlongingDesign/}}.

\subsection{The Prevalence of EPDs across VLOPs}
\begin{figure}[H]
    \centering
  \includegraphics[width= \linewidth]{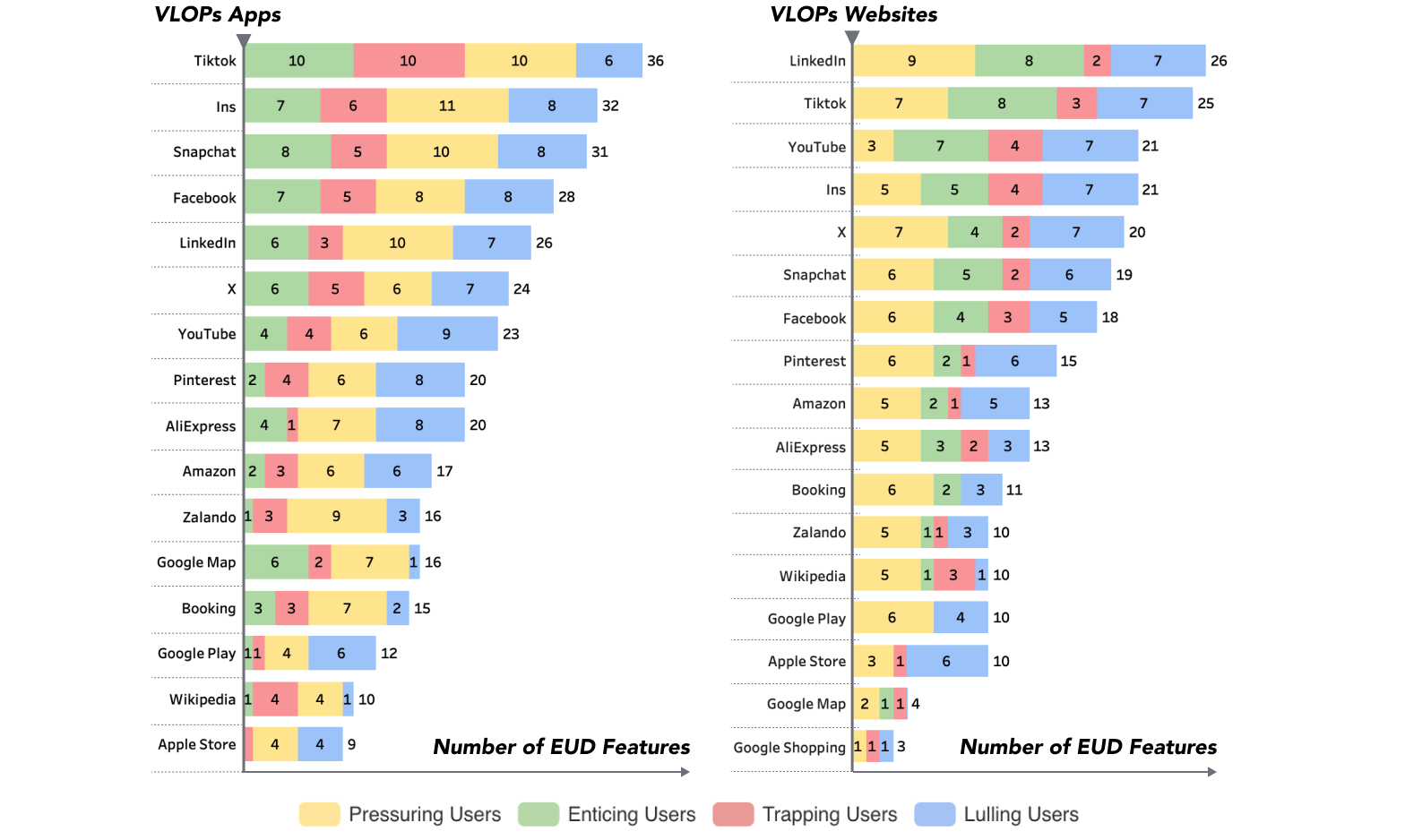}
  \caption{The Prevalence of EPDs by Platforms; app(left), website(right)}
  \label{fig:byplatform}
\end{figure}

We found that EPDs were extremely common on VLOPs, and 100\% of VLOPs contained features that employed one or more of the EPD patterns described above (see Figure~\ref{fig:byplatform}). Across the 16 VLOP app interfaces, we identified 334 instances of features that employ EPD patterns, and across the 17 VLOP website interfaces, we identified 249 instances of features that employ EPD patterns. However, the use of EPDs was not evenly distributed, and for example, we found that TikTok's app interface contains 36 different features that draw on EPD patterns, while the Apple App Store app uses only nine. On average, VLOP apps ($N=16$) used $mean=20.94$ EPD features ($SD=8.10$, $min=9$, $max=36$, $median=20$), with TikTok, Instagram, Snapchat, Facebook, and LinkedIn each using more than 25. The prevalance of EPDs was similar on VLOP websites ($N =17$), where platforms employed an average of $mean=14.65$ features ($SD=6.75$, $min=3$, $max=26$, $median=13$). Website version of LinkedIn, Tiktok, YouTube, Instagram, and X topped the list, each including more than 20 features that employ EPDs.

\begin{figure}[H]
    \centering
  \includegraphics[width= 0.7\linewidth]{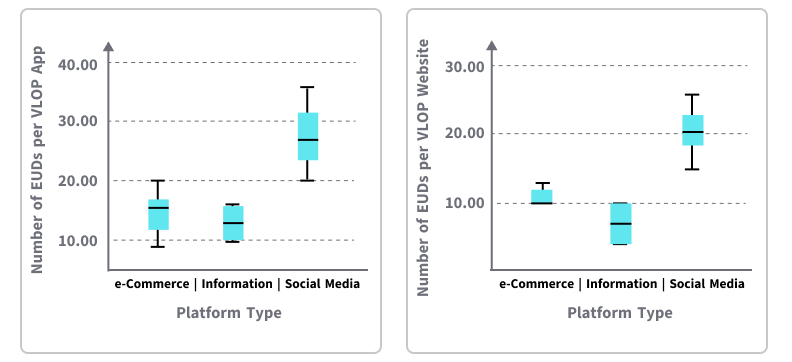}
  \caption{The Prevalence of EPDs by Platform Type(eCommerce, information, social media); app(left), website(right)}
  \label{fig:eudsbytype}
\end{figure}

We examined the prevalence of features that use EPDs by platform type, coding each platform as one of: 1) social media ($N=8$), 2) eCommerce ($N=6$), and 3) information ($N=2$). We used non-parametric tests to compare counts, given the small number of platforms in each group. An independent-samples Kruskal-Wallis test revealed a significant effect of platform-type on number of total EPDs ($\chi^2(2)=11.05$, $p=.004$) on VLOP Apps. Post hoc comparisons revealed that social media VLOPs have significantly more EPDs ($mean=27.5$, $sd=5.3$) than either eCommerce VLOPs ($mean=14.8$, $sd=3.9$, $\chi^2(2)=-2.96$, $p=.003$) or information VLOPs ($mean=13$, $sd=4.2$, $\chi^2(2)=-2.31$, $p=.021$, see Figure~\ref{fig:eudsbytype}).
The independent-samples Kruskal-Wallis test for VLOP websites also revealed a significant effect of platform-type on number of total EPDs (\(\chi^2(2) = 8.86\), \(p = .012\)). Post hoc comparisons revealed that social media VLOPs have significantly more EPDs (\(mean = 20.63\), \(sd = 3.66\)) than either eCommerce VLOPs (\(mean = 10.00\), \(sd = 3.32\), \(\chi^2(2) = -2.65\), \(p = .0082\)) or information VLOPs (\(mean = 7.00\), \(sd = 4.24\), \(\chi^2(2) = -2.24\), \(p = .025\)), see Figure~\ref{fig:eudsbytype}.

\subsection{The Prevalence of each Type of EPD}
\begin{figure}[H]
    \centering
  \includegraphics[width=0.8\textwidth]{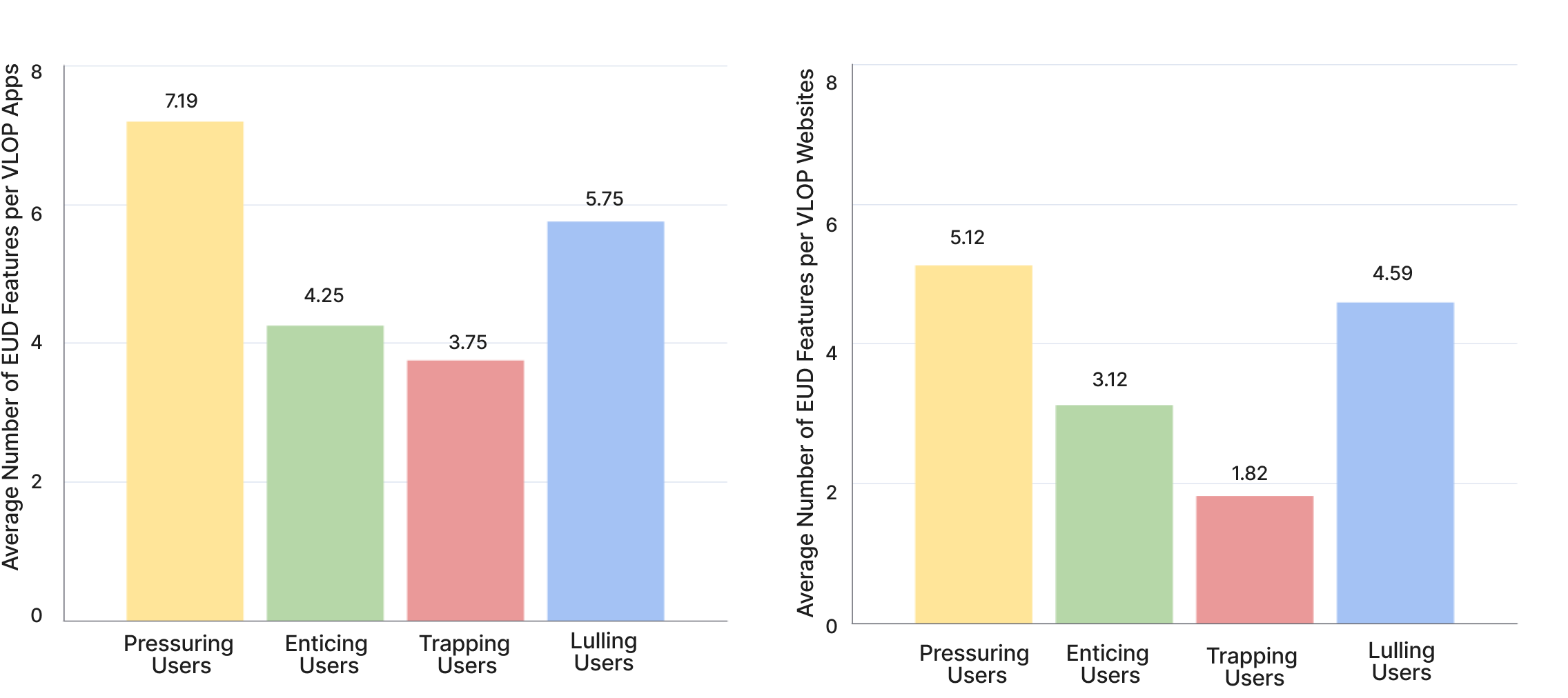}
  \caption{Average Number of EPD Features per platform;  app(left), website(right)}
  \label{fig:anova}
\end{figure}

We found that all four EPD approaches (i.e., pressuring, enticing, trapping, and lulling) were used widely, with all but five VLOPs using all four approaches on both their web and app interfaces (see Figure~\ref{fig:byplatform}). Of the 334 EPD features on VLOP apps, 34.4\% pressure users, 27.5\% lull users, 20.1\% entice users, and 17.9\% trap users. We conducted a repeated measures ANOVA comparing the prevalence of each type of EPD across apps, finding a significant difference in frequency ($F(3)=5.84, p<.001, \eta^2=.226$), with EPDs that pressure users ($mean=7.2, sd=2.29$) being significantly more common than designs that trap ($mean=3.75, sd=2.27$) or lull ($mean=5.75, sd=2.7$) them (see Figure~\ref{fig:anova}). These findings were consistent with the use of EPDs on VLOP websites as well. Of the 249 EPD features on VLOP websites, 34.9\% pressure users, 31.7\% lull users, 21.2\% entice users and, and 12.4\% trap users. A repeated measures ANOVA comparing the prevalence of each type of EPD across websites revealed a significant difference in frequency ($F(3)=8.20, p<.001, \eta^2=.278$) with EPDs that pressure users ($mean=5.12, sd=1.96$) being significantly more common than EPDs that trap them ($mean=1.82, sd=1.24$) or entice ($mean=3.12, sd=2.71$) them.

\begin{figure}[H]
    \centering
  \includegraphics[width=1\linewidth]{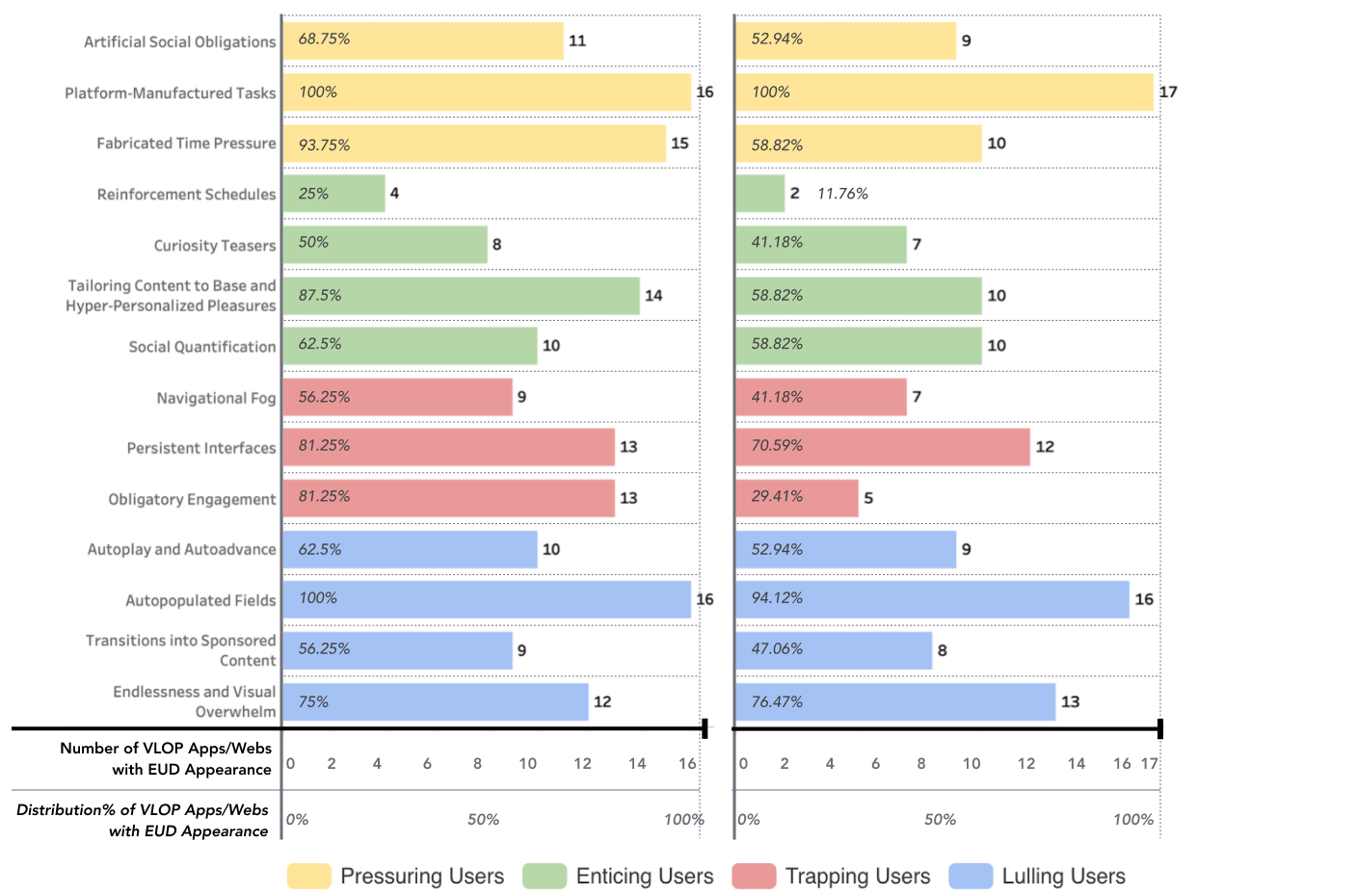}
  \caption{EPD Pattern Occurrences on VLOPs; app(left), website(right)}
  \label{fig:distribution}
\end{figure}

\begin{figure}[H]
    \centering
    \includegraphics[width=0.7\linewidth]{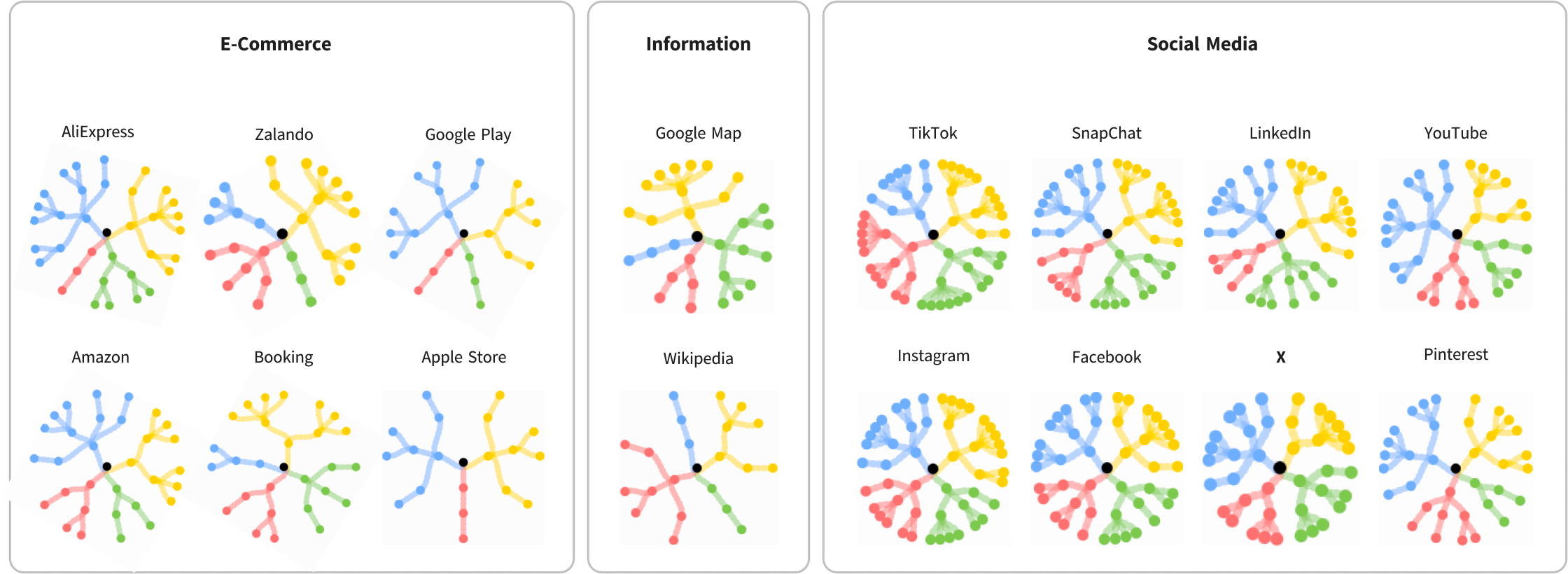}
    \caption{Tree diagrams of the EPDs on each VLOP app. Each node represents a different EPD, and each color represents a different strategy (yellow: pressuring users; green: enticing users; red: trapping users; blue: lulling users). As shown here, eCommerce platforms and information platforms are visibly sparser than social media platforms, which use twice as many EPDs. Interactive versions of these visualizations that allow for exploring the EPDs used by each app can be found on the Engagement-Prolonging Design website\protect\footnotemark.}
    \label{fig:treeMaps}
\end{figure}

We also looked at the prevalence of each of these four strategies across the three types of VLOPs (social media, eCommerce, information). We found that social media platforms (apps: $mean=19.00, sd=7.55$; websites: $mean=8.38, sd=2.13$) generally use pressuring users and lulling users (apps: $mean=29.00, sd=14.01$; websites: $mean=7.63, sd=0.92$) as their main strategies. In contrast, e-commerce platforms primarily use pressuring users (apps: $mean=13.00, sd=7.55$; websites: $mean=6.17, sd=1.94$) and lulling users (apps: $mean=16.00, sd=14.01$; websites: $mean=4.83, sd=2.23$). Information platforms tend to use trapping users (apps: $mean=2.00, sd=7.21$; websites: $mean=3.00, sd=1.41$) and pressuring users (apps: $mean=4.00, sd=7.55$; websites: $mean=5.50, sd=2.12$), though they generally use fewer engagement strategies overall compared to social media and e-commerce platforms. Figure~\ref{fig:treeMaps} shows the tree diagram of each VLOP's usage of EPDs.

\footnotetext{\url{https://engagementprolongingdesign.github.io/EngagementProlongingDesign/}}

\section{Results Part 3: Interface Vignettes Combining EPDs}
Here, we complement our data describing EPDs and their prevalence with illustrations of their use in context. We provide three different vignettes that characterize the holistic experience of encountering EPDs as a VLOP user and show how multiple EPDs can work together, each compounding the effects of the others. 

\subsection{Vignette 1: Creating Chain Reactions}
We observed that VLOPs often funnel users through multiple screens that each incorporate one or more EPDs. These chain reactions create a behavioral funnel that guides users step-by-step through a series of interactions, with each encouraging prolonged engagement by nudging them into the step that follows.

\begin{figure}[H]
    \centering
  \includegraphics[width=1\linewidth]{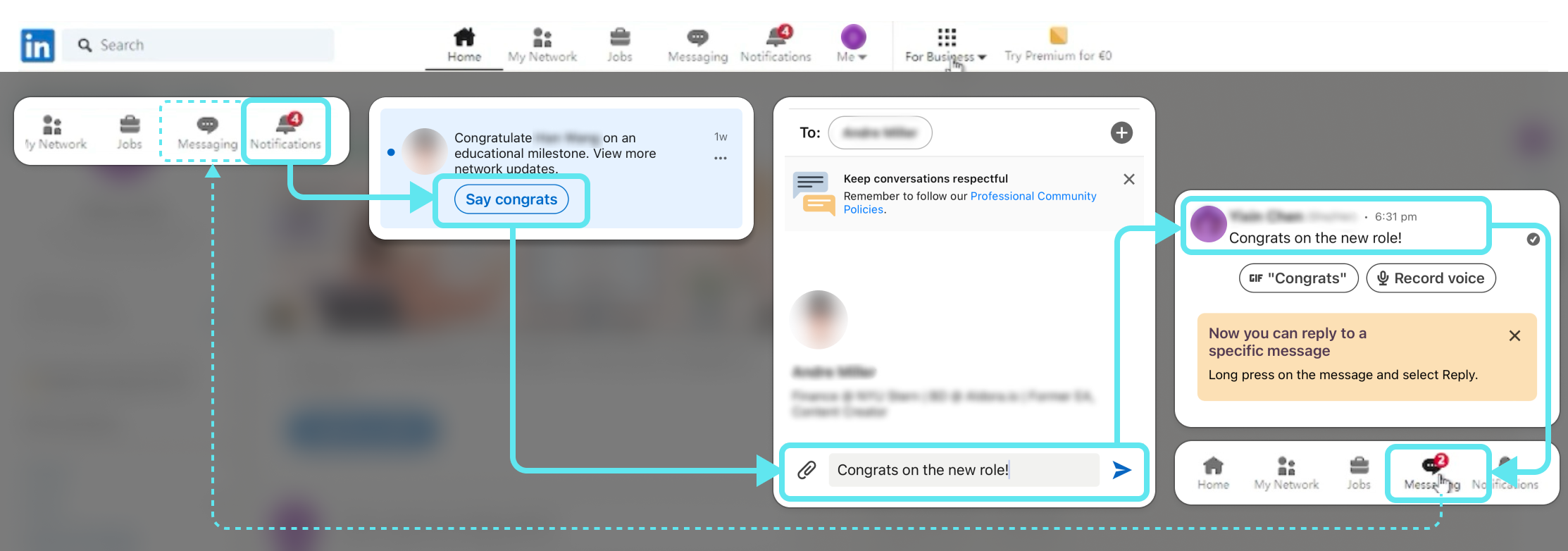}
  \caption{Creating Chain Reactions: LinkedIn creates chain reaction by having one EPD feature lead to another.}
  \label{fig:chainReaction}
\end{figure}

For example (see storyboard in Figure~\ref{fig:chainReaction}), LinkedIn creates an unread notification badge as a visual cue to trigger \textit{Obligatory Engagement}. This creates a subtle sense of urgency or discomfort, prompting the user to click on the badge to clear it. Upon clicking the badge, the user encounters a prompt encouraging them to congratulate a friend on achieving a milestone, creating a subtle \textit{Artificial Social Obligation} without overt pressure. When the user clicks on the prompt encouraging them to ``Say congrats,'' they are taken to a chat interface with an \textit{Autopopulated Field} supplying the text to send, requiring only one click to send a congratulatory message. Sending this message then triggers EPDs for the recipient: the recipient receives a social activity alert in the form of an in-app badge, push notification, or email, informing them of the message. This, in turn, creates another incomplete task that requires action to click the button, and generate social pressure to respond.

With this set of EPD features, the platform creates a complete feedback loop:  even though both features---a notification badge and prompts to congratulate a friend---rely on mechanisms that subtly pressure the user, their combination also creates a nuanced experience of lulling the users into a state where they are more likely to follow a pre-designed path of actions. This gently guides them into extended engagement while subduing the feeling of being pressured or manipulated to stay on the platform. Low-friction designs like autopopulated congratulatory text further lull the user into extended engagement. And this chain reaction moves across the network of connected individuals, with one user's actions triggering EPDs in another user's interface.

\subsection{Vignette 2: Bombarding Users via Multiple Lines of Attack}
In other cases, VLOPs employ multiple EPDs separately and simultaneously, bombarding the user with many parallel attempts to hijack their attention. For example, on YouTube’s main page, features based on lulling mechanisms, like infinite scroll and autoplay, provide a continuous flow of content, keeping users passively engaged without requiring active effort. Features based on pressuring mechanisms, such as unread badges, subtly push users to resolve notifications at the same time. Features based on enticing mechanisms, like provocative thumbnails, instantly grab attention, even as the user scrolls through options or clears notifications. Features based on trapping mechanisms, like picture-in-picture playback, keep users connected to content even as they maintain other threads of engagement. These overlapping features work together to cumulatively create an environment that aggressively pursue the user's attention.

%Our interview results further confirm this effect. During the interviews, we asked participants about their experiences when encountering multiple EUD features on one page. We found that when many elements of an interface work together to engage the user, they tend to have a stronger sense of being manipulated or become more aware of this combination of designs aimed at extending engagement. As P4 noted, 'I think users are kind of bombarded with all these different extended use designs at the same time, whether they're in their peripheral vision or their main focus on the screen.' P3 echoed this sentiment, stating that it was easy to find combined EUD features on one page: 'When I say it's easy to find, it's because I have a strong sense that it manipulates me to engage --- my immediate impression is that the entire page feels very engaging.

\begin{figure}[H]
    \centering
  \includegraphics[width=1\linewidth]{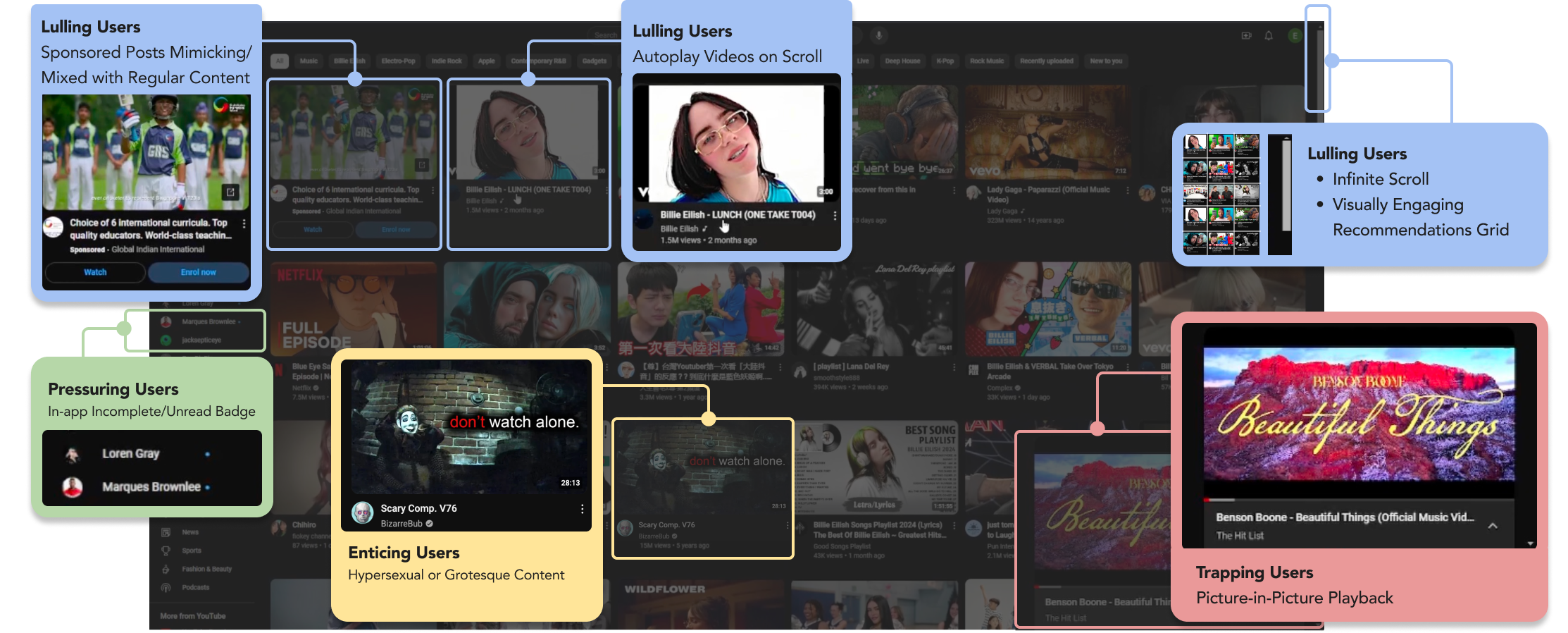}
  \caption{Bombarding the user: YouTube displays a range of EPDs representing all four strategies on the main page.}
  \label{fig:interface}
\end{figure}

\subsection{Vignette 3: Fusing EPDs}
\begin{figure}[H]
    \centering
  \includegraphics[width=1\linewidth]{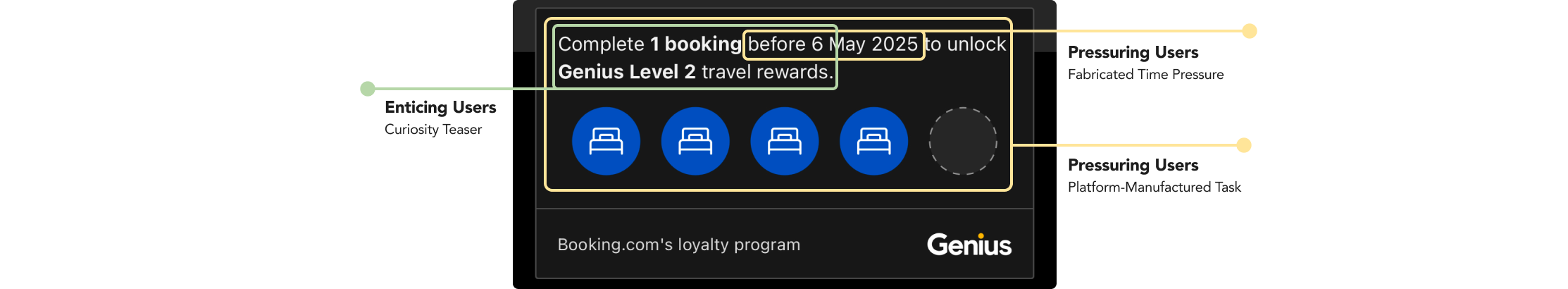}
  \caption{Fusing EPDs: Booking.com creates a \textit{Platform-Manufactured Task} with \textit{Fabricated Time Pressure}, employing multiple forms of pressure through a single feature.}
  \label{fig:fusion}
\end{figure}

Finally, we saw that VLOPs often merge two or more EPD patterns into a single UI element to enhance its potential for extending user engagement. For example, Booking.com presents the user with a dialog encouraging them to complete a series of bookings with the platform within a certain time frame (see Figure~\ref{fig:fusion}). In doing so, they create a \textit{Platform-Manufactured Task}, giving the user a to-do list of bookings, and they exert \textit{Fabricated Time Pressure}, giving the user a time constraint which adds urgency to their engagement. The user's progress is represented by five circles, which break down the broader goal (completing five bookings) into smaller, manageable tasks (one booking at a time). By merging these two EPDs, a single UI element can leverage multiple approaches at once to increasing the user's engagement.

\section{Discussion}\label{sec:Discussion}

% The Ubiquity of EUDs
% Axes/quadrants
% Design Vision
% The Challenge of Gray Patterns
% Lack of Teen-Specific Experience
%

\subsection{The Ubiquity and Aggressiveness of EPDs}
We found that EPDs are ubiquitous across VLOPs, which serve hundreds of millions of users worldwide. These design categories have been invented by platforms to encourage prolonged use and re-engagement, and thus contribute to displacement of other pursuits and lead users to blame themselves for excessive time spent with engaging interfaces online. We found that platforms: 1) pressure users into engaging by creating artificial social obligations and task lists, 2) entice users by leveraging reward- and pleasure-mediated mechanisms that often involve curiosity teasing and tailoring, 3) trap users through disorientating interface elements that cause confusion, block navigation, and inhibit decision-making, and 4) lull users via automated, low-friction designs that reduce user self-determination by automating decision-making, overloading the UI with heuristic-driven stimuli, and encouraging mindlessness. 

These categories of manipulation work together to both guide users toward decisions that extend their use and away from decisions that end their use. Enticement designs attempt to make users feel excited to stay engaged while pressureful designs make them feel bad about leaving. Trapping designs make leaving difficult, while lulling designs make staying easy. VLOPs leverage these strategies in combination, linking them together to form chain reactions, bombarding the user with a multitude of competing EPDs, and fusing multiple EPDs into single features. Together, these forces create a current that carries the user along in one direction, steering them away from disengagement decisions while simultaneously funneling them into engagement ones. 

\subsection{A Conceptual Model of Engagement-Prolonging Design}
This taxonomy builds upon several pre-existing taxonomies that have described dark patterns, attention capture, and designs that promote compulsive use~\cite{flayelle2023taxonomy, monge2023defining, radesky2022prevalence, gray2024ontology} and situates them on VLOPs, the most commonly used platforms with the greatest global reach. Its novel contributions to this academic field include its combination of theory-driven and empirically driven data to create a taxonomy that describes the motives behind design, rather than only describing designs phenomenologically. In other words, while dark pattern typologies describing specific interface designs such as ``confirmshaming'' may come and go, we aimed to describe the underlying purpose and psychological mechanisms by which users pay attention to digital products, use them for extended periods of time, return after non-use, or feel compelled to re-engage. %Therefore, we propose that our taxonomy will be easier to apply to a range of digital products and have more longevity as designs change. Such longevity will be particularly important as generative artificial intelligence informs novel user interface and marketing designs.

\begin{figure}[htbp]
    \centering
  \includegraphics [width=1\linewidth]{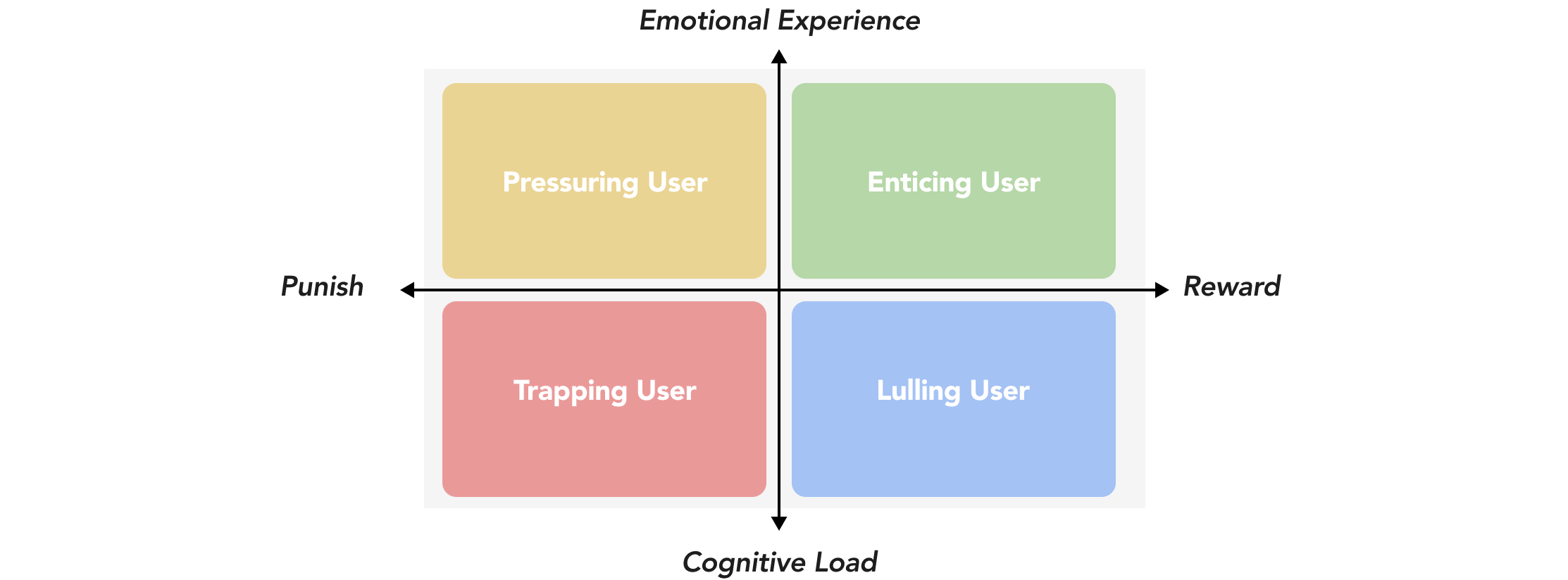}
  \caption{A conceptual model of the common strategies underlying EPDs. Some EPDs manipulate emotions (e.g., manufacturing feelings of guilt, anxiety, or stress by \textit{pressuring} users; manufacturing feelings of curiosity, pleasure, or popularity by \textit{enticing} users). Some EPDs manipulate cognitive load (e.g., making navigation difficult by \textit{trapping} users; making navigation simple by \textit{lulling} users). These designs work to reward users emotionally and cognitively for continuing to engage and punishing users emotionally and cognitively for attempting to disengage.}
  \label{fig:eudMatrix}
\end{figure}

Our findings show that EPDs operate on two levels (see Figure~\ref{fig:eudMatrix}). First, they leverage users' \textit{emotional responses}, manufacturing guilt, anxiety, curiosity, and pleasure in ways that encourage engagement and discourage disengagement. Second, they manipulate \textit{cognitive load}, making engagement decisions frictionless and disengagement decisions frictionful. In both cases, EPDs offer rewards for behaviors that extend engagement and punish behaviors that end it. This model offers some potential insight into future regulatory strategies; for example, punishing designs that invoke negative emotional experiences and increase cognitive load may be better suited to regulation than those that reward users for extended engagement. The parallel structure that we repeatedly encountered---wherein engagement flows (e.g., signing up for a service) lulled users into continuing, and disengagement flows (e.g., canceling a service) trapped users such that they struggled to make forward progress---also suggests a need for regulation. Requiring that engagement and disengagement experiences have equal levels of lulling and trapping would support users in making more autonomous decisions about their time and attention.

Perhaps not coincidentally, these four strategies align with the four phases of the ``Hook'' product-development model, a product design framework popularized by the book of the same name~\cite{eyal2014hooked}, that promises to help companies build habit-forming products. The four phases of the Hook model include: Trigger (creating external and internal triggers to engage, which aligns with the practice of enticing Users), Action (making engagement simple, which aligns with the practice of lulling Users), Variable Reward (encouraging the user to chase unpredictable rewards, which aligns with the practices of pressuring and enticing users), and Investment (requiring users to put effort into the platform, which aligns with the practice of trapping users). It is perhaps no coincidence that the practices we found VLOPs to be employing to encourage extended use fit neatly with practitioner guidance teaching designers how to do the same.

\subsection{The Challenge of ``Gray Patterns''}
One challenge we observed when examining these patterns is that they are often agnostic to user wellbeing rather than outright hostile toward it. In some cases, a user may appreciate a nudge to wish a friend a happy birthday, the convenience of autoplay, or the rush of maintaining a ``streak.'' These designs may, at times, offer value or cement habits users would like to maintain. This makes the prospect of banning or regulating them far more challenging. It also adds complexity for product teams who want to design features that align user- and company-interests.

However, prior work documents that it is deeply problematic to assume that users' engagement behaviors always reflect their preferences~\cite{richards2024against}. Using extended use as a measure of users' interest in a product is confounded by the very fact that EPDs intend to manufacture extended use itself. And it is clear that attention-economy incentives drive the proliferation of EPDs: in our sample, social media platforms that depend on user attention employed twice as many EPDs as platforms that rely on other business models. This suggests both a need for regulation and landscape that defies it, leaving regulators with the challenge of curtailing overly aggressive or invasive EPDs while leaving space for products to engage their users in ways they find pleasurable and meaningful. 

Historically, dark patterns have been identified by their explicit deception, but engagement has avoided regulation because of the slipperiness of its definition between metric, ideology, and harmful practice \cite{richards2024against}. In the case of EPDs, users may be deceived primarily by their own lack of awareness of their susceptibility to these patterns \cite{luguri2021shining, m2020towards}. Our results show that EPDs are pervasive and, if effective, have the potential to dramatically reshape how people spend their time. This suggests a need for regulations that go beyond straightforward cases of deception and put bounds on companies' ability to subtly manipulate users via gray patterns (that are concerning but not quite ``dark'').

\subsection{A Lack of Autonomy-Supporting Experiences For Teens}
Perhaps one of the most notable aspects of our dataset of the EPDs a teen would encounter on VLOPs was the \textit{lack} of teen-specific experiences. The designs platforms employed when a minor user was on the platform were consistent with those an adult might expect to encounter. Children and teens will be particularly susceptible to EPDs due to higher sensitivity to social feedback from peers, which will make pressure more effective in prolonging engagement. The behavior of children and teens is also more easily shaped through reinforcement schedules and pleasure-seeking, which makes them more susceptible to enticement. With less developed critical thinking skills, meta-cognition, and self-monitoring, children and teens are more likely to follow the flow of navigation cues and be trapped or lulled by deceptive designs. It is notable that these design categories frequently appear in the digital products that children are most likely to use through school hours and overnight---namely, social media, video-sharing, and gaming platforms~\cite{monge2023defining}.

Here, we envision several possible design approaches that address the problematic aspects of EPDs (see Table~\ref{tab:designs}). Because EPDs vary in their degree of manipulation, age-inappropriateness, and ethics, we propose solutions that range from prohibiting certain designs on platforms used by minors, to providing low-friction options to turn off a given design pattern. These counter-designs serve the purpose of increasing user agency and transparency to child and teen users about their options. Shifting design patterns from promoting extended use to supporting self-determination and disengagement would support youth wellbeing by reducing excessive time online and improving families’ sense of self-efficacy around technology.

\begin{table}[ht]
\centering
\setlength{\extrarowheight}{2pt}
\begin{tabularx}{\textwidth}{|c|X|}
\hline
\textbf{Mechanism} & \textbf{What designs that do not exploit the mechanism might look like} \\
\hline
\textbf{Pressuring Users} & 
\begin{itemize}
    \item Reduce time pressures: provide options for retaining ephemeral content longer (e.g., allow Instagram stories to last longer).
    \item Reduce artificial social pressures by limiting requests to connect and de-personalizing UI elements, making social nudges "opt-in," or facilitating turning off notifications within 1 click.
    \item Eliminate platform-manufactured tasks.
    \item Eliminate notifications about ads and sponsored content.
    \item Prohibit anthropomorphic interfaces at decision points for minors.
\end{itemize} \\
\hline
\textbf{Enticing Users} & 
\begin{itemize}
    \item Provide options to remove reinforcement schedules in products used by minors (e.g., 1-click turning off of social quantification markers, daily rewards, Snap Streaks, etc.).
    \item Increase human review of content and thumbnails to remove clickbait images that are deceptive or exploit heuristics like violence or sex.
    \item Eliminate curiosity teasers in products used by minors and reduce their use for adults (e.g., 1-click option to turn them off).
    \item Increase transparency about A/B testing or automated tailoring of content (e.g., "because you liked...").
\end{itemize} \\
\hline
\textbf{Trapping Users} & 
\begin{itemize}
    \item Design clearer UIs that support user disengagement (e.g., clear stoppage points, inclusion of clocks, setting time goals for platform usage, clear "X" buttons to exit tasks).
    \item Ensure language is simple and understandable for minors.
    \item Make profile changes easier to apply broadly with minimal clicks.
    \item Provide clear options for canceling or ending accounts.
    \item Include 1-click toggles to disable persistent interfaces permanently.
\end{itemize} \\
\hline
\textbf{Lulling Users} & 
\begin{itemize}
    \item Include 1-click options to turn off autoplay (autoplay off by default for minors).
    \item Provide 1-click options to disable curated or algorithmic feeds.
    \item Disable autofilled searches for minors' accounts.
    \item Increase human review of thumbnails to remove manipulative visuals that exploit cognitive biases.
\end{itemize} \\
\hline
\end{tabularx}
\caption{Potential design approaches that combat EPDs, support users' autonomy, and protect minors.}
\label{tab:designs}
\end{table}

\subsection{Limitations}
There were several limitations to our ability to review VLOP interfaces. First, we did not create paid membership accounts, which prevented us from capturing features like Amazon Prime’s one-click-buy and other experiences designed for premium users. Second, we only examined Android and Windows systems, as these are the most-used operating systems in Europe, and VLOPs on iOS systems have slight differences in their design. Third, our short-term observation method cannot capture EPDs that arise over prolonged use, nor does it fully emulate the experience of a teenager interacting with the platform. We simulated the usage patterns of a teenage user by creating youth personas and accounts, which we found to yield valuable insights, but future work with actual teenage users would be a valuable complement to the data presented here. 

Finally, this research was conducted outside of a corporate context. EPDs that are not readily observable (e.g., data tracking, personalization, and backend designs) would not have been possible for us to detect, and many of the proposed interventions would be difficult or impossible to implement without the cooperation of VLOPs themselves. Greater transparency and collaboration between researchers and industry stakeholders are crucial for building on the findings we present.

\section{Conclusion}\label{sec:Conclusion}
We conducted an inductive-deductive thematic analysis of all VLOPs recognized by the European Union to identify and characterize the EPDs these platforms use with teen users. We then distilled these into a taxonomy and found that designs clustered into four categories reflecting the strategies that underlie them: 1) pressuring, 2) enticing, 3) trapping, and 4) lulling users into engaging. These approaches manipulate users' emotional experience (i.e., pressuring, enticing) and users' cognitive load (i.e., lulling, trapping). They work together to reward users emotionally and cognitively for continuing to engage and to punish users emotionally and cognitively for attempting to disengage. 

We then conducted a structured content analysis, applying our taxonomy to all 17 VLOPs. This revealed that EPDs are pervasive, with 583 features employing these approaches. Although all VLOPs used at least some EPDs, social media platforms used twice as many as other VLOPs. In the wild, EPDs do not operate in isolation, but are chained together to trigger one another, act simultaneously on the user, and combine into single features, thus turning what might, in isolation, be subtle nudges into a cacophony of aggressive attention-grabbing gimmicks.

\section*{Selection and Participation of Children}
No children participated in this work.

% \begin{acks}
% thanks.
% \end{acks}

\bibliographystyle{ACM-Reference-Format}
\bibliography{references}

\end{document}